\newif\ifplainstyle
\plainstyletrue
\plainstylefalse

\newif\ifjhepstyle
\jhepstyletrue
\jhepstylefalse

\newif\ifprstyle
\prstyletrue

\ifprstyle
	\documentclass[twocolumn,nofootinbib]{revtex4-1}
\else
	\documentclass[11pt,a4paper]{article}
\fi

\ifjhepstyle
	\usepackage{jheppub}
	\usepackage{amsfonts}
	\usepackage{verbatim}
	\usepackage{float}
	\usepackage{color}
	\usepackage{array}
	\newcolumntype{C}[1]{>{\centering\arraybackslash$}p{#1}<{$}}
	\makeatletter
	\def\@fpheader{\phantom{Prepared for submission to JHEP}}
	\makeatother
\else	
 	\ifprstyle
		\usepackage{verbatim}
		\usepackage{amsmath,amsfonts,amssymb}
		\usepackage[colorlinks=true
                	,urlcolor=blue
                	,anchorcolor=blue
                	,citecolor=blue
                	,filecolor=blue
                	,linkcolor=blue
                	,menucolor=blue
                	]{hyperref}
	\else
            	\usepackage{verbatim}
            	\usepackage{cite}
            	\usepackage{setspace}
            	\usepackage[top=2.5cm, bottom=2.75cm, left=2.5cm, right=2.5cm]{geometry}
            	\usepackage{amsmath,amsfonts,amssymb}
            	\usepackage[colorlinks=true
            	,urlcolor=blue
            	,anchorcolor=blue
            	,citecolor=blue
            	,filecolor=blue
            	,linkcolor=blue
            	,menucolor=blue
            	,linktoc=page
            	]{hyperref}
            	\usepackage{float}
            	\restylefloat{table}
            	
            	\numberwithin{equation}{section}
	\fi
\fi

\allowdisplaybreaks

\usepackage{tikz}
\usepackage{pgfplots}
\usepackage{bbold}
\pgfplotsset{compat=1.14}

\newcommand{\ThisIsTheTitle}{Bounds on CFTs with \texorpdfstring{${\cal W}_3$}{W3} algebras and \texorpdfstring{AdS$_3$}{AdS3} higher spin theories} 
\newcommand{\ThisIsAuthorOne}{Luis Apolo}
\newcommand{\ThisIsEmailOne}{luis.apolo@fysik.su.se}
\newcommand{\ThisIsTheAffiliation}{Department of Physics \& The Oskar Klein Centre, Stockholm University, \\
 AlbaNova University Centre, SE-106 91 Stockholm, Sweden}
\newcommand{\TheseAreTheKeywords}{}

\newcommand{\ThisIsTheAbstract}{The scaling dimension of the first excited state in two-dimensional conformal field theories (CFTs) satisfies a universal upper bound. Using the modular bootstrap, we extend this result to CFTs with ${\cal W}_3$ algebras which are generically dual to higher spin theories in AdS$_3$. Assuming unitarity and modular invariance, we show that the conformal weights $h$, $\bar{h}$ of the lightest charged state satisfy $h  < c/12 + {\cal O}(1)$ and $\bar{h}  < \bar{c}/12 + {\cal O}(1)$ in the limit where the central charges $c$, $\bar{c}$ are large. Furthermore, we show that in this limit any consistent CFT with ${\cal W}_3$ currents must contain at least one state whose ${\cal W}_3$ charge $w$ obeys $|w| > 4 |h-c/24| /\sqrt{10 \pi c} + {\cal O}(1)$. We discuss hints on the existence of stronger bounds and comment on the interpretation of our results in the dual higher spin theory.
}

\ifjhepstyle
\title{\ThisIsTheTitle}

\author{\ThisIsAuthorOne}

\affiliation{\ThisIsTheAffiliation}

\emailAdd{\ThisIsEmailOne}

\abstract{\ThisIsTheAbstract} 

\fi

\begin{document}

\ifjhepstyle
\maketitle
\flushbottom
\fi

\long\def\symfootnote[#1]#2{\begingroup%
\def\thefootnote{\fnsymbol{footnote}}\footnote[#1]{#2}\endgroup} 

\def\rednote#1{{\color{red} #1}}
\def\bluenote#1{{\color{blue} #1}}

\def\({\left (}
\def\){\right )}
\def\lb{\left [}
\def\rb{\right ]}
\def\lB{\left \{}
\def\rB{\right \}}

\def\Int#1#2{\int \textrm{d}^{#1} x \sqrt{|#2|}}
\def\Bra#1{\left\langle#1\right|} 
\def\Ket#1{\left|#1\right\rangle}
\def\BraKet#1#2{\left\langle#1|#2\right\rangle} 
\def\Vev#1{\left\langle#1\right\rangle}
\def\Vevm#1{\left\langle \Phi |#1| \Phi \right\rangle}\def\bbox{\bar{\Box}}
\def\til#1{\tilde{#1}}
\def\wtil#1{\widetilde{#1}}
\def\ph#1{\phantom{#1}}

\def\ra{\rightarrow}
\def\la{\leftarrow}
\def\lra{\leftrightarrow}
\def\p{\partial}
\def\diff{\mathrm{d}}

\def\sinh{\mathrm{sinh}}
\def\cosh{\mathrm{cosh}}
\def\tanh{\mathrm{tanh}}
\def\coth{\mathrm{coth}}
\def\sech{\mathrm{sech}}
\def\csch{\mathrm{csch}}

\def\a{\alpha}
\def\b{\beta}
\def\g{\gamma}
\def\d{\delta}
\def\e{\epsilon}
\def\ve{\varepsilon}
\def\k{\kappa}
\def\l{\lambda}
\def\n{\nabla}
\def\om{\omega}
\def\s{\sigma}
\def\t{\theta}
\def\z{\zeta}
\def\vp{\varphi}

\def\ss{\Sigma}
\def\dd{\Delta}
\def\gg{\Gamma}
\def\ll{\Lambda}
\def\tt{\Theta}

\def\A{{\cal A}}
\def\B{{\cal B}}
\def\cE{{\cal E}}
\def\D{{\cal D}}
\def\F{{\cal F}}
\def\H{{\cal H}}
\def\I{{\cal I}}
\def\J{{\cal J}}
\def\K{{\cal K}}
\def\L{{\cal L}}
\def\N{{\cal N}}
\def\O{{\cal O}}
\def\P{{\cal P}}
\def\cS{{\cal S}}
\def\W{{\cal W}}
\def\X{{\cal X}}
\def\Z{{\cal Z}}

\def\mfa{\mathfrak{a}}
\def\mfb{\mathfrak{b}}
\def\mfc{\mathfrak{c}}
\def\mfd{\mathfrak{d}}

\def\we{\wedge}
\def\re{\textrm{Re}}

\def\tilw{\tilde{w}}
\def\tile{\tilde{e}}

\def\zz{\bar z}
\def\xx{\bar x}
\def\xp{x^{+}}
\def\xm{x^{-}}

\def\VirU1{\mathrm{Vir}\otimes\hat{\mathrm{U}}(1)}
\def\VirSL2R{\mathrm{Vir}\otimes\widehat{\mathrm{SL}}(2,\mathbb{R})}
\def\U1{\hat{\mathrm{U}}(1)}
\def\SL2R{\widehat{\mathrm{SL}}(2,\mathbb{R})}
\def\sl2r{\mathrm{SL}(2,\mathbb{R})}
\def\by{\mathrm{BY}}

\def\RR{\mathbb{R}}

\def\tr{\mathrm{Tr}}
\def\bnabla{\overline{\nabla}}

\def\sint{\int_{\ss}}
\def\dsint{\int_{\p\ss}}
\def\hint{\int_{H}}

\newcommand{\eq}[1]{\begin{align}#1\end{align}}
\newcommand{\eqst}[1]{\begin{align*}#1\end{align*}}
\newcommand{\eqsp}[1]{\begin{equation}\begin{split}#1\end{split}\end{equation}}

\newcommand{\absq}[1]{{\textstyle\sqrt{\left |#1\right |}}}

\ifprstyle
\title{\ThisIsTheTitle}

\author{\ThisIsAuthorOne}
\email{\ThisIsEmailOne}


\affiliation{\ThisIsTheAffiliation}


\begin{abstract}
\ThisIsTheAbstract
\end{abstract}


\maketitle

\fi

\ifplainstyle
\begin{titlepage}
\begin{center}

\ph{.}

\vskip 4 cm

{\Large \bf \ThisIsTheTitle}

\vskip 1 cm

\renewcommand*{\thefootnote}{\fnsymbol{footnote}}

{{\ThisIsAuthorOne}\footnote{\ThisIsEmailOne} }

\renewcommand*{\thefootnote}{\arabic{footnote}}

\setcounter{footnote}{0}

\vskip .75 cm

{\em \ThisIsTheAffiliation}

\end{center}

\vskip 1.25 cm

\begin{abstract}
\noindent \ThisIsTheAbstract
\end{abstract}

\end{titlepage}

\newpage

\fi

\ifplainstyle
\tableofcontents
\noindent\hrulefill
\bigskip
\fi

\section{Introduction}
\label{se:intro}

The modular bootstrap is a powerful tool used to constrain the spectrum of two-dimensional conformal field theories (CFTs). Like the conformal bootstrap, which relies on conformal and crossing symmetry of correlation functions~\cite{Belavin:1984vu}, the modular bootstrap also relies on symmetry, namely invariance under large diffeomorphisms of CFTs defined on the torus.\footnote{A generalization of the modular bootstrap to higher dimensional CFTs is not known (for generalizations of modular invariance see refs.~\cite{Cardy:1991kr,Belin:2016yll,Shaghoulian:2016gol}.) In contrast, the conformal bootstrap has been applied to CFTs in three and four dimensions, see e.g.~\cite{Rattazzi:2008pe,Poland:2011ey,ElShowk:2012ht}.} 

A remarkable consequence of modular invariance is Cardy's formula~\cite{Cardy:1986ie}. There, modular S-transformations, $\tau \ra -1/\tau$ where $\tau$ is the complex structure of the torus, are used to relate the low ($-i\tau \ra \infty$) to the high ($-i \tau \ra 0$) temperature limits of the partition function. As a result, the density of states at high temperature and large central charge is universal for any unitary, modular-invariant CFT.

Modular S-transformations yield further universal constraints on CFTs at large central charge~\cite{Hellerman:2009bu,Hellerman:2010qd,Keller:2012mr,Friedan:2013cba,Qualls:2013eha,Qualls:2014oea,Qualls:2015bta,Kim:2015oca,Benjamin:2016fhe,Collier:2016cls}. These follow from the observation that, at the fixed point $\tau = i$, there is an \emph{infinite} number of constraints on the partition function $Z(\tau,\bar{\tau})$ of modular invariant CFTs~\cite{Hellerman:2009bu},
  \eq{
  (\tau \p_{\tau} )^{n} (\bar{\tau} \p_{\bar{\tau}})^{m}  Z(\tau,\bar{\tau}) \Big |_{\tau=i} = 0, \qquad n + m = \mathrm{odd.} \label{hellerman}
  }
In the presence of a chemical potential $\mu$, these constraints have been generalized to~\cite{Benjamin:2016fhe}
  \eq{
   (\tau \p_{\tau} )^{n} (\bar{\tau} \p_{\bar{\tau}})^{m} (\p_{\mu})^{l}  \Big [Z ({\cal S} x_i) - {\cal S} Z (x_i) \Big ]_{{\tau = i, \mu = 0}} = 0, \label{bootstrap}
  }
where $\cS \O$ denotes the S-transformed value of $\O$ and $x_i = \{\tau, \bar{\tau}, \mu\}$ stands for the arguments of the partition function.

In this paper we constrain the conformal weights of the lightest charged state in CFTs with $\W_3$ algebras and prove the existence of a state whose mass-to-charge ratio obeys a universal upper bound. The $\W_3$ algebra is a nonlinear extension of the Virasoro algebra which, in the semiclassical, large central charge limit, reproduces the asymptotic symmetries of $SL(3)\times SL(3)$ higher spin theories in AdS$_3$~\cite{Campoleoni:2010zq}.\footnote{In the context of the AdS/CFT correspondence~\cite{Maldacena:1997re,Gubser:1998bc,Witten:1998qj}, CFTs with $W_N \times W_N$ algebras are \emph{generically} dual to $SL(N) \times SL(N)$ higher spin theories in AdS$_3$, which feature fields of spin $s = 2, \dots, N$~\cite{Henneaux:2010xg,Campoleoni:2010zq,Campoleoni:2011hg}. Concrete examples of the duality are known for $N \ra \infty$~\cite{Gaberdiel:2010pz} (see~\cite{Gaberdiel:2012uj} for a review).} Assuming unitarity and modular invariance, we find that the conformal weights $h$, $\bar{h}$ of the first excited state carrying ${\W_3}$ charge satisfy
  \eq{
  h < \frac{c}{12} + \O(1), \qquad \bar{h} < \frac{\bar{c}}{12} + \O(1), \label{chiralbound}
  }
in the limit where the central charges $c$, $\bar{c}$ are large. The bound~\eqref{chiralbound} holds for CFTs with either one or two copies of the $\W_3$ algebra. In particular, for a left-right symmetric theory where $c = \bar{c}$, the scaling dimension $\dd$ obeys
  \eq{
  \dd < \frac{c}{6} + \O(1). \label{bound}
  }

The bound~\eqref{bound} is shared by other (nonextremal) CFTs without supersymmetry.\footnote{Known exceptions satisfying stronger bounds are extremal CFTs with a holomorphically factorized partition function~\cite{Witten:2007kt}, and CFTs with $\N=(1,1)$ or $\N = (2,2)$ supersymmetry where the object of interest is the elliptic genus~\cite{Gaberdiel:2008xb,Benjamin:2016fhe}.} Indeed, it is also satisfied by the lightest uncharged state in CFTs with or without extended chiral symmetries~\cite{Hellerman:2009bu,Friedan:2013cba,Qualls:2013eha,Qualls:2015bta}, and by charged states in CFTs with abelian chiral symmetries~\cite{Benjamin:2016fhe}. In all of these cases, it is not possible to rule out the existence of stronger bounds inaccessible to analyses based on eqs. \eqref{hellerman} and \eqref{bootstrap} with a finite number of derivatives. Indeed, ref.~\cite{Collier:2016cls} foregoes this assumption by considering a large number of derivatives (extrapolated to infinity) before taking the large-c limit; these limits do not seem to commute, and numerical evidence is presented for a stronger bound, $\dd < \a c + \O(1)$ where $\frac{1}{12} < \a < \frac{1}{9}$. One of the motivations for the present paper was to confirm whether the non-abelian and nonlinear structure of the $\W_3$ algebra is constraining enough to reproduce similar bounds, even when the number of derivatives in eq.~\eqref{bootstrap} is small. While our analysis shows that this is not the case, we discuss hints that suggest the existence of a stronger bound where $\dd < \frac{c}{12} + \O(1)$. 

Turning to the dual theory, we note that $SL(3) \times SL(3)$ higher spin theories admit black hole solutions carrying $\W_3$ charge~\cite{Gutperle:2011kf} (see~\cite{Ammon:2012wc} for a review). These higher spin black holes obey an extremality bound which, in terms of CFT quantities, reads~\cite{Gutperle:2011kf,Bunster:2014mua,Banados:2015tft}
  \eq{
  \dd_{BH} \ge \frac{c}{12} + \frac{3}{2} \( \frac{5}{3} c\, w^2 \)^{1/3},
  }
where the $\W_3$ charge $w$ and the central charge $c$ are assumed to be equal in the left and right-handed sectors. Thus, in the semiclassical, large central charge limit, the CFT bound~\eqref{bound} is larger than the lightest charged higher spin black holes.  

Using the modular bootstrap, we also prove the existence of a state whose $\W_3$ charge $w$ is bounded, in the large-$c$ limit, by
  \eq{
  |w| > \sqrt{\frac{8}{5\pi c}}\, \left | h - \tfrac{c}{24} \right | + {\cal O}(1). \label{chargebound}
  }
Thus, if $h \ne \tfrac{c}{24}$, the CFT must feature a state with $|w| > N^2 \, c^{1/2}$ for some nonvanishing number $N$. In particular, assuming a left-right symmetric CFT, eq.~\eqref{chargebound} implies the existence of a state in the dual higher spin theory whose mass-to-charge ratio satisfies\footnote{The case $\dd = \tfrac{c}{12}$ must be handled with care. If $h \ne \tfrac{c}{24}$, then $Q$ is still bounded by terms of $\O(\sqrt{c})$, as in eq.~\eqref{chargebound}. Otherwise if $h = \tfrac{c}{24}$, then $Q$ is bounded by $Q > \sqrt{\tfrac{E_4(i)}{900\pi}} + \O(1/c)$, cf.~eq.~\eqref{wboundlargec2}.}
  \eq{
  \left | \frac{\dd - \tfrac{c}{12}}{Q} \right | < \sqrt{\frac{5\pi c}{2}} + \O \Big (c^{-1/2} \Big ).\label{masstochargeratio}
  }
In eq.~\eqref{masstochargeratio} we have defined $Q = \tfrac{1}{2} (|w| + |\overline{w}|)$ where $w$ and $\overline{w}$ denote the $\W_3$ charges of the dual higher spin theory. The existence of a state satisfying eq.~\eqref{masstochargeratio} is reminiscent of the weak gravity conjecture, which features a similar bound~\cite{ArkaniHamed:2006dz}.\footnote{For generalizations of the weak gravity conjecture to AdS$_d$ spacetimes see e.g.~\cite{Nakayama:2015hga}. In particular, for related studies in AdS$_3$ that focus on the structure of the dual CFT see refs.~\cite{Benjamin:2016fhe,Montero:2016tif}.} However, in our case the bound cannot be interpreted as a measure of the relative strength between two forces since the dual higher spin theory contains only one coupling constant, namely $G_N$. It would be interesting to understand the necessity of this bound from the point of view of the dual higher spin theory.

The bounds~\eqref{bound} and~\eqref{chargebound} constrain states in the CFT and, correspondingly, in the dual higher spin theory. However, the latter is deemed inconsistent if it contains light matter fields~\cite{Perlmutter:2016pkf}. There are two exceptions: (i) pure higher spin gravity, an analog of pure three-dimensional gravity, which admits only boundary gravitons, boundary higher spin fields, and black holes; and (ii) higher spin theories with an infinite tower of higher spin fields dual to CFTs with $\W_{\infty}[\l]$ algebras. If pure higher spin gravity exists, the bound~\eqref{bound} constrains the scaling dimension of the lightest black hole microstate, while~\eqref{chargebound} constraints the charge of at least one such microstate. On the other hand, we comment on the obstacles of extending our analysis to CFTs with $\W_{\infty}[\l]$ algebras.


\section{Partition function and modular transformations}
\label{se:partition}

Let us begin by defining the partition function and deriving its modular transformation in CFTs with a chiral $\W_3$ algebra. The symmetries of the theory are described by~\cite{Zamolodchikov:1985wn}
  \eq{
  [\bar{L}_n,\bar{L}_m] &= (n-m) \bar{L}_{n+m} + \frac{\bar{c}}{12} n (n^2 -1) \d_{n+m}, \label{barredvirasoro} \\
  [{L}_n,{L}_m] &= (n-m) {L}_{n+m} + \frac{{c}}{12} n (n^2 -1) \d_{n+m}, \\
  [L_n, W_m] &= (2n - m) W_{n+m},\\
  \begin{split}
  [W_n, W_m] &= \frac{1}{15} (n-m) (n^2 -\tfrac{1}{2} nm + m^2 - 4) L_{n+m} \\
  			+ \xi(n-&m) \ll_{n+m} + \frac{c}{360} n(n^2-1)(n^2-4) \d_{n+m}, 
\end{split} \label{w3algebra}
  }
where $\bar{L}_n$ and $L_n$ represent the modes of the chiral components of the stress-energy tensor $\overline{T}(\bar{z})$ and $T(z)$, $W_n$ denotes the modes of the dimension-3 current $W(z)$, and $\xi$ is given by
  \eq{
  \xi = \frac{16}{22+5c}. \label{xi}
  }
The modes $\ll_n$ in eq.~\eqref{w3algebra} are those of a composite operator $\ll(z)$ which is responsible for the nonlinearity of the algebra,
  \eqsp{
  \ll_{n} =& \sum_{p=1}^{\infty} \(L_{-p} L_{n+p} + L_{n-p} L_{p}\) + L_n L_0 \\
  & + \frac{1}{10} (n+2)(1-3n) L_n.
  }

We are interested in constraining the conformal weights $h$, $\bar{h}$ of states carrying $\W_3$ charge, i.e.~states $\Ket{\psi} = \Ket{h,\bar{h},w}$ such that
  \eqsp{
  L_0 \Ket{\psi} &= h \Ket{\psi}, \qquad W_0 \Ket{\psi} = w \Ket{\psi}, \\
  \bar{L}_0 \Ket{\psi} &= \bar{h}\Ket{\psi}.
  }
Therefore, we focus on the canonical partition function with an additional chemical potential $\mu$ for the dimension-3 current
  \eq{
  Z(\tau,\bar{\tau},\mu) & = \tr\,(q^{L_0 - k} \bar{q}^{\bar{L}_0 - \bar{k}} y^{W_0}), \label{partitionbare} \\
  & = \sum_{h,\bar{h},w} d_{h\bar{h}w} \,q^{h - k} \bar{q}^{\bar{h} - \bar{k}} y^{w}. \label{partitionfunction}
  }
In eqs.~\eqref{partitionbare} and~\eqref{partitionfunction} we have assumed a discrete spectrum, $q = e^{2\pi i \tau}$, $y = e^{2\pi i \mu}$, and we have defined $k = \tfrac{c}{24}$, $\bar{k} = \tfrac{\bar{c}}{24}$ for convenience. 

Besides the $\W_3$ algebra, the additional requirements made on the CFT are \emph{unitarity} and \emph{modular invariance}. The former restricts the $d_{h\bar{h}w}$ coefficients in eq.~\eqref{partitionfunction} to be positive, while the latter leads to a modular invariant partition function when $\mu = 0$, i.e.~
  \eq{
  Z \(\tau',\bar{\tau}', 0\) = Z(\tau, \bar{\tau}, 0),\label{zerothorder}
  }
where $\tau'$ is given by
  \eq{
  \tau' = \frac{\a \tau + \b}{\g \tau + \d}, \label{modularp}
  }
with $\{\a,\b,\g,\d\} \in \mathbb{Z}$ and $\a\d-\b\g=1$. Once the chemical potential is turned on, the partition function is no longer modular invariant. While a closed-form expression for the transformed partition function is not known, it is possible to determine this transformation perturbatively around $\mu = 0$~\cite{Gaberdiel:2012yb,Iles:2014gra}. In particular, note that the constraints given in eq.~\eqref{bootstrap} are also expanded around this point and that, to recover the bound in eq.~\eqref{chiralbound}, we will only need terms at most quadratic in $\mu$. The modular-transformed partition function thus reads
  \eq{
  Z(\tau', \bar{\tau}', \mu') = \sum_{n=0} \frac{(2\pi i)^n}{(\g \tau + \d)^{3n}} \frac{\mu^n}{n!} \Vev{W_0^n}_{\tau', \bar{\tau}'} , \label{perturbativepartition}
  }
where $\Vev{\O}_{\tau, \bar{\tau}} = \tr\,(\O \,q^{L_0 - k} \bar{q}^{\bar{L}_0 - \bar{k}})$ denotes the torus one-point function of $\O$, and $\mu'$ is given by~\cite{Li:2013rsa}\footnote{The transformation~\eqref{modularmu} is also featured in ref.~\cite{Kaneko:1995abc} in the context of generalized Jacobi Theta functions.}
  \eq{
  \mu' = \frac{\mu}{(\g \tau + \d)^{3}}. \label{modularmu}
  }

Let us now determine the transformation of the partition function up to quadratic order in $\mu$. To zeroth order, the partition function is modular invariant, cf.~eq.~\eqref{zerothorder}. This is also true at linear order where we have
  \eq{
  \Vev{W_0}_{\tau', \bar{\tau}'} & = (\g \tau + \d)^3 \Vev{W_0}_{\tau, \bar{\tau}}. \label{firstorder}
  }
This equation follows from the fact that (i) only the zero mode contributes to the one-point function of primary fields on the torus; and (ii) for any primary field $\O_{h,\bar{h}} (z)$ with weights $h$ and $\bar{h}$, the torus one-point function transforms as
  \eq{
  \Vev{\O_{h,\bar{h}}}_{\tau', \bar{\tau}'} = (\g \tau + \d)^{h} (\g \bar{\tau} + \d)^{\bar{h}} \Vev{\O_{h,\bar{h}}}_{\tau, \bar{\tau}}. \label{primarytransformation}
  }

Next, we consider the transformation of the partition function at quadratic order in $\mu$. Following refs.~\cite{Iles:2013jha,Iles:2014gra}, we first construct a composite primary field $\P(z)$ such that (see Appendix~\ref{ap:a} for details) 
  \eq{
  \Vev{\P}_{\tau, \bar{\tau}} = \Vev{W_0^2}_{\tau, \bar{\tau}} + \sum_{n=0}^{3} c_n (\tau) \Vev{L_0^n}_{\tau, \bar{\tau}},
  }
where $L_0^0$ is the identity operator. We then note that the modular transformations of $\Vev{\P}_{\tau,\bar{\tau}}$ and $\Vev{L_0^n}_{\tau, \bar{\tau}}$ are easily found: the former is given by eq.~\eqref{primarytransformation} while the latter follows directly from eq.~\eqref{zerothorder}. In this way, the modular transformation of $\Vev{W_0^2}_{\tau, \bar{\tau}}$ is determined entirely from the symmetry algebra and the modular invariance of the CFT. Deferring details of the derivation to Appendix~\ref{ap:a} we find
  \eqsp{
   \Vev{W_0^2}_{\tau', \bar{\tau}'} = (\g \tau + \d)^6 \bigg [ &\Vev{W_0^2}_{\tau, \bar{\tau}} - \frac{i}{\pi}\frac{\g}{\g \tau + \d} G_{\tau, \bar{\tau}} \bigg ], \label{secondorder}
  }
where the function $G_{\tau,\bar{\tau}}$ is given by
  \eqsp{
  G_{\tau,\bar{\tau}} = \frac{\xi}{12} \bigg \{ & \frac{c}{240} \Big[2 E_4(\tau) + 20 E_2(\tau) + 5 c \Big ] Z(\tau,\bar{\tau},0) \\
    & \,\,\, - \Big [ 2 E_2(\tau) + c \Big] \Vev{L_0}_{\tau, \bar{\tau}} + \Vev{L_0^2}_{\tau, \bar{\tau}} \bigg \},\label{G}
  }
and $E_ 2(\tau)$, $E_4(\tau)$ denote the Eisenstein series,
  \eq{
  E_2(\tau) = 1 - 24 \sum_{n=1}^{\infty} \frac{n q^n}{1- q^n},\\
  E_4(\tau) = 1 +240 \sum_{n=1}^{\infty} \frac{n^3 q^n}{1- q^n}.
  }
In particular, note that eqs.~\eqref{secondorder} and~\eqref{G} agree with, and generalize, the results of ref.~\cite{Iles:2014gra}.

We conclude that, up to quadratic order in $\mu$, the modular transformation of the partition function is given by
  \eq{
  Z(\tau', \bar{\tau}',\mu') = Z(\tau,\bar{\tau},\mu) + 2 \pi i \mu^2 \frac{\g}{\g \tau + \d} G_{\tau,\bar{\tau}} + \dots \label{modulart}
  }
Equation~\eqref{modulart} generalizes the corresponding equation in CFTs with a chiral $U(1)$ algebra. There, the transformation of the partition function $Z_{U(1)}(\tau,\bar{\tau},\mu)$ is known to all orders~\cite{Kraus:2006nb}\footnote{The partition function $Z_{U(1)}(\tau,\bar{\tau},\mu)$ is given by eq.~\eqref{partitionbare} with $W_0$ the zero mode of a dimension-1 current and $\mu$ a chemical potential whose modular transformation is given by $\mu' = \frac{\mu}{\g \tau + \d}$.}
  \eq{
  Z_{U(1)}(\tau', \bar{\tau}',\mu') = e^{2\pi i \mu^2 \frac{\g}{\g \tau + \d}} Z_{U(1)}(\tau,\bar{\tau},\mu),\label{U1}
  }
and its expansion around $\mu=0$ reproduces eq.~\eqref{modulart} with $G_{\tau,\bar{\tau}} \ra Z_{U(1)}(\tau,\bar{\tau},0)$. In contrast, the fact that $G_{\tau,\bar{\tau}}$ in eq.~\eqref{G} depends not only on $Z(\tau,\bar{\tau},0)$, but also on $\Vev{L_0}_{\tau,\bar{\tau}}$ and $\Vev{L_0^2}_{\tau,\bar{\tau}}$, is a consequence of the non-abelian and nonlinear structure of the $\W_3$ algebra.


\section{Bounds on charged states}
\label{se:bound}

We now have the necessary ingredients to derive bounds on the charged states of CFTs with $\W_3$ algebras. The latter follow from the constraints imposed by invariance of the CFT under modular S-transformations, cf.~eq.~\eqref{bootstrap}.\footnote{Recall that modular S-transformations correspond to $\a = \d = 0$ and $\beta = -\g = -1$ in eqs.~\eqref{modularp}, \eqref{modularmu}, and \eqref{modulart}.} These constraints may be written as
  \eq{
  \sum_{h,\bar{h},w} d_{h\bar{h}w}\, \L F(h,\bar{h},w) \Big |_{\tau = i, \mu = 0}  = 0, \label{bootstrap2}
  }
where $\L$ is a differential operator given by
  \eq{
  \L = \sum_{n,m,l} c_{nml}  (\tau \p_{\tau} )^{n} (\bar{\tau} \p_{\bar{\tau}})^{m} (\p_{\mu})^{l}, \label{Loperator}
  }
and $F(h,\bar{h},w)$ reads
  \eqsp{
  \!\!\!F(h,\bar{h},w) \!= & e^{-\frac{2\pi i}{\tau} (h-k) + \frac{2 \pi i}{\bar{\tau}} (\bar{h} - \bar{k}) + \frac{2 \pi i \mu}{\tau^3}w} - \tfrac{2 \pi i \mu^2}{\tau} G_{\tau,\bar{\tau}}\\
  &-  e^{2\pi i \tau( h-k) -2\pi i \bar{\tau} (\bar{h}-\bar{k}) + 2 \pi i \mu w}. \label{F1}
  } 
For convenience, in the remainder of the text $\L F(h,\bar{h},w)$ is evaluated at $\tau = i$ and $\mu = 0$.

Unitarity restricts the $d_{h\bar{h}w}$ coefficients in eq.~\eqref{bootstrap2} to be positive. Thus, the modular constraints are satisfied by any nontrivial operator $\L$ if the corresponding $\L F(h, \bar{h}, w)$ function admits both positive and negative values.  In particular, assume that $h^*$ and $\bar{h}^*$ exist such that $\L F(h, \bar{h}, w)$ is positive for all $h > h^*$ and $\bar{h} > \bar{h}^*$, and is negative otherwise. Then, eq.~\eqref{bootstrap2} is satisfied only if states with $h < h^*$ and $\bar{h} < \bar{h}^*$ are included in the spectrum. As a consequence, the conformal weights of the first excited state cannot be larger than $h^*$ and $\bar{h}^*$~\cite{Hellerman:2009bu}. 

\begin{figure}[!h]
\begin{center}
\includegraphics{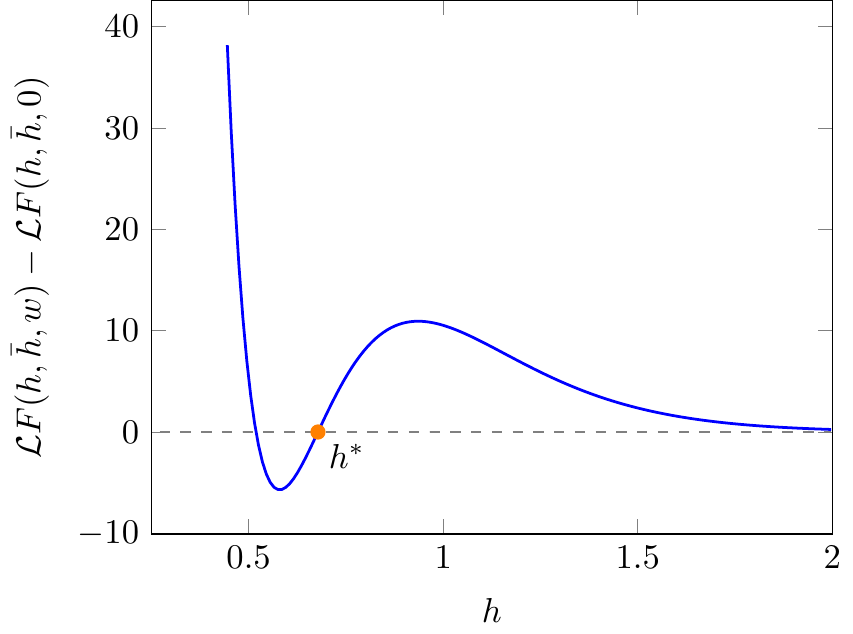}
\end{center}
\caption{Behavior of the $\L F(h,\bar{h},w)$ $-$ $\L F(h,\bar{h},0)$ functions considered in Sections~\ref{se:boundh} and~\ref{se:boundhbar} for $\bar{h} = \bar{k}$ and $k = \frac{1}{24}$. When conditions~\eqref{condition1} and~\eqref{condition2} hold, the lightest charged state must satisfy $h < h^*$.}
\label{lonelyfigure}
\end{figure}

The condition on $\L F(h,\bar{h},w)$ considered above must be modified slightly in order to constrain the conformal weights of charged states. Following ref.~\cite{Benjamin:2016fhe}, our goal is to find operators $\L$ such that, 
  \eq{
  &\L F(0, 0, 0) > 0,  \label{condition1} \tag{i} \\
  & \L F(h, \bar{h}, 0) \ge 0, \label{condition2} \tag{ii} \\
  \begin{split}
  &\L F(h, \bar{h}, w)-\L F(h, \bar{h}, 0) > 0 \quad \textrm{if} \quad h > h^* \quad \textrm{or}\\
  &\quad \bar{h} > \bar{h}^*, \label{condition3}
  \end{split} \tag{iii}
  }
for the smallest possible $h^*$ and $\bar{h}^*$. The first and second conditions tell us that there is at least one positive contribution to eq.~\eqref{bootstrap2}, that of the vacuum, which can only be cancelled by the contribution of charged states. On the other hand, condition~\eqref{condition3} tells us that the charged states necessary to satisfy eq.~\eqref{bootstrap2} obey an upper bound on their conformal weights, namely $h < h^*$ and $\bar{h} < \bar{h}^*$. The behavior of the $\L F(h, \bar{h}, w)-\L F(h, \bar{h}, 0)$ functions considered in Sections~\ref{se:boundh} and~\ref{se:boundhbar} is illustrated in fig.~\ref{lonelyfigure}.

\vspace{10pt}

\subsection{Bound on \texorpdfstring{$h$}{h}}
\label{se:boundh}

Let us first consider the bound on the conformal weight $h$ of the lightest charged state. Note that it is not difficult to find an operator $\L$ satisfying condition~\eqref{condition3}. The simplest example is
  \eq{
  \L =  -(\tau \p_{\tau} )^{3} (\p_{\mu})^{2}, \label{L1}
  }
for which $h^*$, the would-be bound if conditions~\eqref{condition1} and~\eqref{condition2} were satisfied, is given by
  \eq{
  h^* = \frac{c}{24} + \frac{3}{2 \pi}. \label{wouldbehbound}
  }
For this choice of $\L$ the function $\L F(h, \bar{h}, 0)$ is not positive definite, however. The goal then is to find additional contributions to $\L$ such that conditions~\eqref{condition1} and~\eqref{condition2} hold. This inevitably increases the value of $h^*$ given in eq.~\eqref{wouldbehbound} and weakens the bound to the value reported below.

Restricting $\L$ to contain up to 5 derivatives with respect to $\tau$ we find
  \eqsp{
  \!\!\!\L  = & \frac{9\xi}{4} (\tau \p_{\tau} )^{5} + a_{1} (\tau \p_{\tau})+ a_{2} (\p_{\mu})^{2}  + a_3 (\tau \p_{\tau} ) (\p_{\mu})^{2}  \\
  &      + a_{4} (\tau \p_{\tau} )^{2} (\p_{\mu})^{2}, \label{L2}
  }
where we recall that $\xi = \tfrac{8}{11+60k}$, cf.~eq.~\eqref{xi}. The sign of $(\tau \p_{\tau} )^{5}$ in eq.~\eqref{L2} guarantees that $\L F(h, \bar{h},0)$ is positive for large $h$. On the other hand, the $a_{n}$ coefficients are unwieldy functions of the central charge $c$ that render $\L F(h, \bar{h},0)$ positive for all $h$. Their values are given explicitly in Appendix~\ref{ap:b}. The corresponding $\L F(h,\bar{h}, w)$ function can be written as
  \eq{
  \L F(h,\bar{h}, w) &= e^{-2\pi (h+\bar{h}-k - \bar{k} )} \big [ A(h) + w^2 B(h) \big ], \label{LF1}
  }
where possible choices for $A(h)$ and $B(h)$ are given by
  \eq{
  A(h) &= 144 \,\xi \pi^5 \bigg [h (h-k)^2 \Big (h - \tfrac{5+ 4\pi k}{2\pi} \Big )^2 + A_0 \label{A1} \bigg ], \\
  B(h) & = \frac{4 \pi^3}{5} \big (b_2 h^2  + b_1 h + b_0 \big), \label{B1}
  }
and the $b_n$ and $A_0$ coefficients read
  \eq{
  b_2 =& 360 \pi^3 k, \\
  b_1 =& -180 \pi^2 k (7 + 4 \pi k), \\
  \begin{split}
  b_0 = & -675 - 540 \pi k - 2 \pi^3 k E_4(i) (5+3k) , \label{b0}\\
  A_0 =& \frac{2 \pi ^2 k}{675} \Big \{ 675 + 2 \pi  k \big [ 720 + 90 \pi  k (7 + 2 \pi  k)\\
  & - \pi ^2  (20 - 3 k) E_4(i) \big] \Big\} E_4(i) .
  \end{split}
  }

The $a_n$ coefficients in eq.~\eqref{L1} are determined by demanding the strongest possible bound on the conformal weight of charged states at leading order in $c$.\footnote{The $a_n$ coefficients, and correspondingly the functions $A(h)$ and $B(h)$, are not unique and other alternatives exist that lead to the same large-$c$ bound when the number of derivatives in eq.~\eqref{L1} is small, i.e.~of the same order considered in eq.~\eqref{L2}.} While the function $A(h)$ is manifestly positive, $B(h)$ admits two real roots for positive values of the central charge. In particular, the function $\L F(h,\bar{h},w) - \L F(h,\bar{h},0)$ shares the behavior illustrated in fig.~\ref{lonelyfigure} and condition~\eqref{condition3} is satisfied with $h^*$ given by\footnote{The largest root of eq.~\eqref{B1} yields a bound valid for all $c$. The expression is cumbersome so we only report its value at large $c$.}
  \eq{
  h^* &= \frac{c}{12} + \frac{7}{2\pi} + \frac{E_4(i)}{120} + \O(1/c), \label{hstar}
  }
where the Eisenstein series $E_4(\tau)$ at $\tau = i$ evaluates to
  \eq{
  E_4(i) = \frac{\gg \(\tfrac{1}{4}\)^8}{(2\pi)^6}  \sim 0.49. \label{E4}
  }
Thus, unitarity and modular invariance imply that the lightest charged state in CFTs with $\W_3$ algebras satisfies
  \eq{
  h < \frac{c}{12} + \O(1), \label{hbound}
  }
in the limit where the central charge is large.


\subsection{Bound on \texorpdfstring{$\bar{h}$}{bar h}}
\label{se:boundhbar}

We now find an operator $\bar{\L}$ that yields an upper bound on the conformal weight $\bar{h}$ of the lightest charged state. As before, we note that it is not difficult to satisfy condition~\eqref{condition3}. Consider for example,
  \eq{
  \bar{\L} = - (\tau \p_{\tau} ) (\bar{\tau} \p_{\bar{\tau}})^2 (\p_{\mu})^2, \label{barL1}
  }
for which the would-be bound $\bar{h}^*$ is given by
  \eq{
  \bar{h}^* = \frac{\bar{c}}{24} + \frac{2}{\pi}. \label{wouldbebarhbound}
  }
Not surprisingly, the operator~\eqref{barL1} violates conditions~\eqref{condition1} and~\eqref{condition2}. The latter demand the addition of several terms to $\bar{\L}$ which, once again, weaken the would-be bound $\bar{h}^*$ given in eq.~\eqref{wouldbebarhbound} to that derived below.

The operator $\bar{\L}$ satisfying conditions~\eqref{condition1} --~\eqref{condition3} for large enough $c$ and with the fewest number of derivatives is given by
  \eqsp{
  \bar{\L} = & 16 \pi \bar{k} \xi \, (\tau \p_{\tau} ) (\bar{\tau} \p_{\bar{\tau}})^{2} (\p_{\mu})^{2} +  8\, (\bar{\tau} \p_{\bar{\tau}})^{3} (\p_{\mu})^{2} \\
    & + 10 \pi \bar{k} \xi\,(7+12k) (\bar{\tau} \p_{\bar{\tau}})^{2} (\p_{\mu})^{2} + d_1  (\bar{\tau} \p_{\bar{\tau}}) (\p_{\mu})^{2}     \\
    &   + d_2  (\tau \p_{\tau} ) (\p_{\mu})^{2} + d_3 (\p_{\mu})^{2} +  d_4 (\tau \p_{\tau} )^3 (\bar{\tau}\p_{\bar{\tau}})^{2}  
    \\
    &    + d_5 (\tau \p_{\tau} ) (\bar{\tau} \p_{\bar{\tau}})^{2}  +  d_6 (\tau \p_{\tau} )^3 +  d_7  (\tau \p_{\tau} ).  \label{barL2}
    }
Up to normalization, the coefficients of the first three terms in eq.~\eqref{barL2} guarantee that the contribution of the charged and uncharged states to $\bar{\L} F(h, \bar{h},w)$ is positive for large $h$ and $\bar{h}$. The remaining $d_n$ coefficients, whose values are given in Appendix~\ref{ap:b}, depend on both central charges and render $\bar{\L} F(h, \bar{h},0)$ positive for large c (see below). The corresponding $\bar{\L} F(h,\bar{h},w)$ function reads
  \eq{
  \bar{\L} F(h,\bar{h}, w) &= e^{-2\pi (h+\bar{h}-k - \bar{k} )} \big [C(h,\bar{h}) + w^2 D(\bar{h}) \big ], \label{LF2}
  }
where $C(h,\bar{h})$ and $D(\bar{h})$ are given by
  \eq{
  \begin{split}
  &C(h,\bar{h}) = \xi \Big \{ \big [ 1 - 4 \pi (h-k)  \big]^2 + C_0 \Big \} \\ 
  & \qquad\qquad \qquad \qquad \times \Big \{\bar{h} \big [ 3 - 4\pi (\bar{h} - 2 \bar{k}) \big]^2  + 1\Big \},
  \end{split} \label{Cfunction} \\
  \begin{split}
  &D(\bar{h}) = 32 \pi^3 \Big [ 16 \pi^2 \bar{k} \bar{h}^2 - 8 \pi \bar{k} (1+ 4\pi \bar{k}) \bar{h} \\
  & \,\,\qquad\qquad\qquad\qquad\qquad\qquad  -16 \pi \bar{k}^2 - 9 \bar{k} - 1  \Big ].
  \end{split} \label{Dfunction}
  }
The $C_0$ term in eq.~\eqref{Cfunction} is $c$-dependent and given by
  \eq{
  C_0 = \frac{c \pi^2}{90} E_4(i) -1.
  }

The fact that $\bar{\L}$ in eq.~\eqref{barL2} involves twice as many terms as $\L$ in eq.~\eqref{L2} follows from the nontrivial dependence of $(\bar{\tau} \p_{\bar{\tau}})^n (\p_{\mu})^2 F(h,\bar{h},w)$, for any integer $n > 0$, on both $h$ and $\bar{h}$. This can be ultimately traced back to the non-abelian and nonlinear nature of the $\W_3$ algebra, i.e.~to the $h$-dependence of the transformed partition function~\eqref{modulart}. Clearly, the function $C(h,\bar{h})$ is positive and satisfies conditions~\eqref{condition1} and~\eqref{condition2} for $C_0 > 0$, i.e.~for all $c > 90/(\pi^2 E_4(i) ) \sim 18.8$. Furthermore, $D(\bar{h})$ is proportional to $\bar{h}^2$ for large $\bar{h}$ and admits two real zeroes. Hence the function $\bar{\L} F(h,\bar{h},w) - \bar{\L} F(h,\bar{h},0)$ features the dip shown in fig.~\ref{lonelyfigure}, and condition~\eqref{condition3} is satisfied with $\bar{h}^*$ given by
  \eq{
  \bar{h}^* & = \frac{6+\pi \bar{c}+\Big (864\bar{c}^{-1} + 36(10 + \pi \bar{c}) + \pi^2\bar{c}^2  \Big )^{1/2}}{24\pi}, \\  
  \bar{h}^* & = \frac{\bar{c}}{12} + \frac{1}{\pi} + \O(1/\bar{c}).
  }
Thus the conformal weight $\bar{h}$ of the lightest charged state in CFTs with $\W_3$ algebras is bounded by
  \eq{
  \bar{h} < \frac{\bar{c}}{12} + \O(1), \label{barhbound}
  }
in the limit where the central charges $c$ and $\bar{c}$ are large.

We conclude this section by noting that in the presence of an additional $\W_3$ algebra, denoted here by $\overline{\W}_3$, the modular transformation of the partition function may be modified by terms proportional to $\bar{\mu}^2$, the chemical potential for the $\overline{\W}_3$ charge. In this case the function $F(h,\bar{h},w)$ in eq.~\eqref{F1} receives corrections proportional to $\bar{\mu}^2$. The operators $\L$, $\bar{\L}$ in eqs.~\eqref{L1} and~\eqref{barL2} are blind to these modifications, however, since they are evaluated at the self-dual point $\tau = i$, $\mu = \bar{\mu} = 0$. This implies that the bounds~\eqref{hbound} and~\eqref{barhbound} hold also for theories with an additional $\W_3$ algebra, the latter of which is necessary in the duality to $SL(3) \times SL(3)$ higher spin theories discussed in Section~\ref{se:intro}.


\subsection{Bound on $w$}

Let us now use the modular bootstrap to prove the existence of a state whose mass-to-charge ratio is bounded from above by $\sqrt{c}$. In order to achieve this, the prescription given by conditions~\eqref{condition1} --~\eqref{condition3} must be modified in the obvious way. We now look for operators $\L$ such that
  \eq{
  &\L F(0, 0, 0) > 0,  \label{condition1star} \tag{i*} \\
  & \L F(h, \bar{h}, 0) \ge 0, \label{condition2star} \tag{ii*} \\
  \begin{split}
  &\L F(h, \bar{h}, w) > 0 \quad \textrm{if} \quad |w| < w^*. \label{condition3star}
  \end{split} \tag{iii*}
  }
for the largest possible $w^*$. Conditions~\eqref{condition1star} --~\eqref{condition3star} tell us that the constraint imposed by modular invariance on the partition function, namely eq.~\eqref{bootstrap2}, cannot be satisfied unless the spectrum contains \emph{at least one} charged state with $|w| > w^*$.

In principle, one can construct operators $\L$ satisfying conditions~\eqref{condition1star} --~\eqref{condition3star} so that $w^*$ is independent of the conformal weight. This would result in bounds constraining only the $\W_3$ charges of states in the theory. However, we find that this is not possible in CFTs with $\W_3$ algebras if the number of derivatives in $\L$ is small. The obstruction originates from the $\Vev{L_0}$ and $\Vev{L_0^2}$ dependence of the modular-transformed partition function, which we recall is a direct consequence of the nonlinear and non-abelian structure of the $\W_3$ algebra. In particular, when operators $\L$ with up to two derivatives with respect to $\mu$ and up to five derivatives with respect to $\tau$ are considered, it is not possible to derive bounds that constraint only the $\W_3$ charges. This pattern, which seems to extend to higher derivatives in the modular parameter, may be broken by higher derivatives in the chemical potential. In this case one would need to consider the transformation of the partition function to higher orders.

On the other hand, the nonlinear structure of the $\W_3$ algebra allows us to construct operators $\L$ satisfying conditions~\eqref{condition1star} --~\eqref{condition3star} with $w^*$ linear in the conformal weight. An operator with at most two derivatives in $\mu$ satisfying these requirements is given by 
  \eq{
  \L = \xi (\tau \p_{\tau} ) - 2\pi (\p_{\mu})^{2}. \label{Lp1}
  }
The corresponding $\L \F (h,\bar{h},w)$ function reads
  \eq{
  \L \F (h,\bar{h},w) = 16 \pi^3 e^{-2\pi (h + \bar{h} - k - \bar{k})} \Big [H(h) - w^2  \Big ], \label{LF3}
  }
 where $H(h)$ is manifestly positive and given by
   \eq{
    H(h) = \frac{4}{(11+60k)\pi} \bigg[ (h-k)^2 + \frac{E_4(i) k}{60} \bigg ] . \label{H1}
   }
Since eq.~\eqref{LF3} is positive for all $|w| < w^* = \sqrt{H(h)}$, the constraint~\eqref{bootstrap2} cannot be satisfied --- and a contradiction is reached --- unless one or more states exist with $|w| > w^*$. Thus, the spectrum of CFTs with $\W_3$ currents must feature at least one charged state for which
  \eq{
  |w| > \sqrt{\frac{  4(h-k)^2 + \tfrac{E_4(i) k}{15}  }{(11+60k)\pi}}. \label{wbound}
  }

Let us now note that the states of CFTs with $\W_3$ algebras are constrained by unitarity to satisfy~\cite{Mizoguchi:1988vk}
   \eq{
   h^2 \( h - \frac{1}{16} - \frac{3k}{4}  \) \ge \frac{9(11+60k)}{32} w^2.
   }
This implies that, with the exception of uncharged states with $h = 0$, all the other states in the theory must scale linearly with the central charge. In particular, if $h \ne k$ then, in the large central charge limit, the bound~\eqref{wbound} reads
  \eq{
  |w| > \sqrt{\frac{8}{5\pi c}}\, \left | h - \tfrac{c}{24} \right | + {\cal O}\( 1/\sqrt{c} \).\label{wboundlargec}
  }
Thus the CFT must contain at least one state whose ratio between conformal weight and charge, $|h-\tfrac{c}{24}|/|w|$, is bounded from above by $\sqrt{ 5\pi c / 8}$. On the other hand, if $h = k$ the bound~\eqref{wbound} becomes
  \eq{
  |w| &> \sqrt{\frac{E_4(i)}{900\pi}} + \O(1/c) \sim 0.013 + \O(1/c) \label{wboundlargec2}.
  }

We may expect operators $\L$ featuring a larger (but still finite) number of derivatives to exist that yield stronger bounds at finite $c$, but preserve the leading order behavior observed in eq.~\eqref{wboundlargec}. In this context we note that, considering up to five derivatives in $\tau$ and two derivatives in $\mu$, it is possible to find several operators satisfying conditions~\eqref{condition1star} --~\eqref{condition3star} for large enough $c$. These operators contain terms like $(\tau \p_{\tau} )^3 (\p_{\mu})^{2}$ and $(\tau \p_{\tau} )^2 (\p_{\mu})^{2}$ but yield weaker bounds in the large-$c$ limit.


\section{Discussion}
\label{se:discussion}

In a left-right symmetric theory where $c = \bar{c}$, the bound on the lightest charged state may be written as
  \eq{
  \dd = h + \bar{h} < \frac{c}{6} + \O(1). \label{deltabound1}
  }
This bound is shared by the lightest charged and uncharged states in other CFTs without supersymmetry~\cite{Hellerman:2009bu,Friedan:2013cba,Qualls:2013eha,Qualls:2015bta,Benjamin:2016fhe}. The analyses performed in these references are also based on eqs.~\eqref{hellerman} and~\eqref{bootstrap} with either a small or large --- but always bounded --- number of derivatives. Therefore, we expect a more systematic approach, one where the operators $\L$, $\bar{\L}$ contain a greater but still finite number of derivatives, to leave the bound unchanged at leading order in $c$. On the other hand, we do expect such an analysis to yield a stronger bound at subleading order in the central charge.

Let us recall that the operators $\L = -(\tau \p_{\tau})^3 (\p_{\mu})^2$ and $\bar{\L} = - (\tau \p_{\tau}) (\bar{\tau} \p_{\bar{\tau}})^2 (\p_{\mu})^2$ would yield a stronger bound given by, cf.~eqs.~\eqref{wouldbehbound} and~\eqref{wouldbebarhbound},
  \eq{
  \dd^* < \frac{c}{12} + \O(1), \label{wouldbedeltabound}
  }
except that conditions~\eqref{condition1} and~\eqref{condition2} do not hold. Hence, the operators $\L$ and $\bar{\L}$ must contain additional terms that render the $\L F(h,\bar{h},0)$ and $\bar{\L} F(h,\bar{h},0)$ functions positive. If the number of $\p_{\tau}$, $\p_{\bar{\tau}}$ derivatives in these operators is bounded from above, then some of the additional terms \emph{must} be proportional to $(\p_{\mu})^2$ and they are responsible for weakening the would-be bound~\eqref{wouldbedeltabound}. The reason for this is that $(\tau \p_{\tau})^n (\bar{\tau} \p_{\bar{\tau}})^m F(h,\bar{h},w)$ vanishes when $n + m$ is even and this reduces the number of polynomials in $h$ and $\bar{h}$ we can use to make $\L F(h,\bar{h},0)$ and $\bar{\L} F(h,\bar{h},0)$ positive. \emph{A priori}, this is not a problem if the number of $\p_{\tau}$, $\p_{\bar{\tau}}$ derivatives is unbounded, in which case terms proportional to $(\p_{\mu})^2$ may not be necessary. Thus, relaxing the assumption on a finite number of derivatives before taking the large-c limit, as done in ref.~\cite{Collier:2016cls}, may lead to the stronger bound suggested by eq.~\eqref{wouldbedeltabound}. 

We conclude with comments on the generalization of our results to CFTs with $\W_{\infty}[\l]$ algebras, the latter of which feature currents of dimension $3, 4, \dots, \infty$ (see Appendix~\ref{ap:a}). As before, we would like to constrain the conformal weights of the first excited state charged under the dimension-3 current. Therefore, we use the partition function defined in eq.~\eqref{partitionbare}.\footnote{A similar setup was considered in ref.~\cite{Gaberdiel:2012yb} where the high temperature limit of the partition function was shown to match the free energy of higher spin black holes~\cite{Kraus:2011ds}.} The modular-transformed partition function is no longer given by eq.~\eqref{modulart}, however. This is to be expected, as the $[W_n, W_m]$ commutator features contributions from the modes $U_n$ of the dimension-4 current $U(z)$, cf.~eq.~\eqref{w3algebra2}. Following the steps described in Section~\ref{se:partition} and Appendix~\ref{ap:a}, the partition function now transforms as
  \eqsp{
  Z(\tau', \bar{\tau}',\mu') =\,\, & Z(\tau,\bar{\tau},\mu) + 2 \pi i \mu^2 \frac{\g}{\g \tau + \d} \Big [ G_{\tau,\bar{\tau}} \\
  & + 2\Vev{U_0}_{\tau,\bar{\tau}} \Big ] + \O(\mu^3). \label{modulart2}
  }

The $\Vev{U_0}_{\tau,\bar{\tau}}$ term in eq.~\eqref{modulart2} gives an additional contribution to $F(h,\bar{h},w)$ in eq.~\eqref{F1} which is proportional to the charge $u$ corresponding to the dimension-4 current. This term leaves the contribution of the charged states to $\L F(h,\bar{h},w)$ and $\bar{\L} F(h,\bar{h},w)$ in eqs.~\eqref{LF1} and~\eqref{LF2} unchanged, so condition~\eqref{condition3} still holds. In contrast, the contribution of the uncharged states is no longer positive due to terms linear in $u$. Thus, the $\L$ and $\bar{\L}$ operators given in Section~\ref{se:bound} satisfy conditions~\eqref{condition1} --~\eqref{condition3} only in the sector of zero $u$-charge and the bounds derived in the previous section cannot be extended to CFTs with $\W_{\infty}[\l]$ algebras. To derive similar bounds, it may be necessary to extend the transformation of the partition function to terms quartic, or higher, in $\mu$.\footnote{Note that turning on a chemical potential $\nu$ for the dimension-4 current, and considering only terms linear in $\nu$ in the transformation of the partition function~\eqref{modulart2}, does not help in removing the $u$-dependent terms from $\L F(h,\bar{h},w)$ and $\bar{\L} F(h,\bar{h},w)$.} In this case, additional currents beyond the dimension-4 current featured in eq.~\eqref{modulart2} will contribute to $Z(\tau',\bar{\tau}',\mu')$.


\section*{Acknowledgments}

It is a pleasure to thank Nico Wintergerst for helpful discussions and comments on the manuscript. This work was supported by a grant from the Swedish Research Council.


\appendix

\section{Modular transformation of \texorpdfstring{$\Vev{W_0^2}_{\tau, \bar{\tau}}$}{<W0W0>}}
\label{ap:a}

In this Appendix we derive the modular transformation of $\Vev{W_0^2}_{\tau, \bar{\tau}}= \tr(W_0^2 q^{L_0 - k} \bar{q}^{\bar{L}_0 - \bar{k}})$ for CFTs with $\W_N$ algebras. For generality, we consider the $\W_{\infty}[\l]$ algebra which features an infinite number of currents of dimensions $3, 4, \dots, \infty$ in addition to the stress-energy tensor. These algebras are characterized by a free parameter $\l$ such that, when $\l = N$ is an integer greater than two, it is possible to truncate the currents with $h > N$ and reduce the $\W_{\infty}[N]$ algebra to $\W_N$.

To lowest order, the $\W_{\infty}[\l]$ algebra is given by~\cite{Gaberdiel:2012ku,Gaberdiel:2012uj}
  \eq{
  \!\!\![{L}_n,{L}_m] =& (n-m) {L}_{n+m} + \frac{{c}}{12} n (n^2 -1) \d_{n+m},\label{virasoro2} \\
  \!\!\![L_n, W_m] =& (2n - m) W_{n+m}, \\
  \!\!\![L_n, U_m] =& (3n - m) U_{n+m}, \\
  \begin{split}
\! \!\![W_n, W_m]=& 2(n-m) U_{n+m} + \frac{5}{2} \xi N_3 (n-m) \ll_{n+m} \\
  +  & \frac{N_3}{6} (n-m) (n^2 -\! \tfrac{1}{2} nm + m^2 - 4) L_{n+m} \\
 &  + \frac{c}{144} N_3 \, n(n^2-1)(n^2-4) \d_{n+m}, 
\end{split} \label{w3algebra2}
  }
where $U_n$ denotes the modes of the dimension 4 current $U(z)$ and $N_3$ reads
  \eq{
  N_3 = \frac{16}{5} \z^2 (\l^2 - 4). \label{N3}
  }
In eq.~\eqref{N3} $\z$ is an arbitrary normalization that fixes the coefficient of the most singular term in the $W(z) W(w)$ OPE. We choose
  \eq{
  \z = \frac{1}{8}\frac{1}{\l^2 - 4}, \label{normalization}
  }
for which the $[W_n,W_m]$ commutator~\eqref{w3algebra2} reduces to that of the $\W_3$ algebra, cf.~eq.~\eqref{w3algebra}, modulo the contribution from the dimension-4 current $U(z)$.

Following ref.~\cite{Iles:2014gra}, we first find a composite primary field of dimension 6 whose zero mode contains $\Vev{W_0^2}_{\tau, \bar{\tau}}$. This field is given by
  \eqsp{
  \P(z) = & (W W)(z) + e_1 (T (T T))(z) + e_2 (T T)''(z) \\
  & + e_3 (T'T)'(z) + e_4 (T'' T)(z) + e_5 T''''(z) \\
  & + e_6 (T U)(z) + e_7 U''(z), \label{primary}
  }
where $(AB)(z) = \tfrac{1}{2\pi i} \oint_z dw A(w)B(z)$ denotes the normal\,\!-ordered product of $A(z)$ and $B(z)$. The $e_i$ coefficients in eq.~\eqref{primary} are given by
  \eq{
  e_1 &= -160 (22 + 191 c) \vp, \notag \\
  e_2 &= 6(3848 - 126 c + 335 c^2) \vp, \notag \\
  e_3 &= -3(19272 + 9386 c - 215 c^2)\vp, \notag \\
  e_4 &= 15(3784 + 1266 c - 43 c^2)\vp, \notag \\
  e_5 &= -(3688 - 2506 c - 235 c^2  + 25 c^3)\vp, \notag \\
  e_6 &= - \frac{88}{3(24+c)}, \quad a_7 = -\frac{76+5c}{9 (24+c)}, \notag
  }
where $\vp^{-1} =30 (-1+2 c) (22+5c) (68+7c)$. Next, we note that only the zero mode of $\P(z)$ contributes to the torus one-point function. The full expression for $\Vev{\P_0}_{\tau,\bar{\tau}}$ is long and not particularly illuminating, so we illustrate the steps in the derivation of $\Vev{W_0^2}_{\tau',\bar{\tau}'}$ by focusing only on the $(WW)(z)$ operator. Dropping the $\tau, \bar{\tau}$ dependence from $\Vev{\P_0}_{\tau,\bar{\tau}}$ for convenience we find
  \eqsp{
  \!\Vev{\P_0} \!=& \Vev{W_0^2}+ 2  {\sum_{p=1}^{\infty}}  \Vev{W_{-p}W_{p}} + 12 \,\xi {\sum_{p=1}^{\infty}}  \Vev{L_{-p} L_{p}} \\
  &+ \tfrac{1}{5} (7+6\, \xi) \Vev{L_0} + 6\, \xi \Vev{L_0^2} + 12 \Vev{U_0} + \dots \label{zeromodes}
  }
where we recall that $\xi = \tfrac{16}{22+5c}$ and $\dots$ denotes the zero-mode contributions of the other operators in $\P(z)$. Using the cyclic property of the trace and the symmetry algebra~\eqref{virasoro2} --~\eqref{w3algebra2}, the sums over modes in eq.~\eqref{zeromodes} read\footnote{In more detail, for any mode $\O_p$ such that $[L_0, \O_p] = - p \,\O_p$ we have $\Vev{\O_{-p}\O_{p}}_{\tau,\bar{\tau}} = q^p \Vev{\O_{p}\O_{-p}}_{\tau,\bar{\tau}} =  \tfrac{q^p}{1-q^p} \Vev{[\O_{p}, \O_{-p}]}_{\tau,\bar{\tau}}$.}
  \eq{
  {\textstyle \sum_{p}} \Vev{W_{-p}W_{p}} =& 4 \s_1 \Vev{U_0}  + \tfrac{1}{6} \big [2 \s_3 - (2+c)\xi\s_1 \big ] \Vev{L_0} \notag \\
  & + 2\, \xi \s_1 \Vev{L_0^2} + 4 \,\xi \s_1 {\textstyle \sum_{p}} \Vev{L_{-p} L_{p}} \\
  & + \tfrac{c}{360} (4 \s_1 - 5 \s_3 + \s_5) \Vev{\mathbb{1}}, \notag \\
  {\textstyle \sum_{p}}   \Vev{L_{-p} L_{p}} \,\,=& 2 \s_1 \Vev{L_0} + \tfrac{c}{12} (\s_3 - \s_1) \Vev{\mathbb{1}},
  }
where $\mathbb{1}$ is the identity operator, i.e.~$\Vev{\mathbb{1}} = Z(\tau,\bar{\tau},0)$, and $\s_{2n-1}$ are functions of $\tau$ given by
  \eq{
  \s_{2n-1} = \sum_{p=1}^{\infty} \frac{p^{2n-1} \,q^p}{1-q^p}, \qquad n \ge 1.
  }

The other operators in $\P(z)$ yield similar contributions so that $\Vev{W_0^2}$ in eq.~\eqref{zeromodes} depends only on 
  \eq{
  \Big\{\Vev{\P_0}, \Vev{L_0^n}, \Vev{U_0}, \Vev{L_0 U_0}, \Vev{\mathbb{1}}, \s_{2n-1} \Big | n = 1,2,3 \Big \} \label{pterms}
  }
The terms~\eqref{pterms} are functions of $\tau$ and $\bar{\tau}$ whose modular transformations are easily determined. Indeed, since both $\P(z)$ and $U(z)$ are primary fields, the transformations of $\Vev{\P_0}$ and $\Vev{U_0}$ follow directly from eq.~\eqref{primarytransformation}. Furthermore, we can write the $\Vev{L_0^n \O}$ correlator, where $\O = \{\mathbb{1}, U_0 \}$, as
  \eq{
  \Vev{L_0^n \O} = \sum_{m=0}^n \binom{n}{m} \bigg ( \frac{c}{24}\bigg )^{n - m} \bigg( \frac{1}{2\pi i}\bigg)^m \p_{\tau}^m \Vev{\O}. \label{L0correlator}
  }
Thus, the modular transformation of eq.~\eqref{L0correlator} is determined from
  \eq{
  \p_{\tau'} = (\g \tau + \d)^2 \, \p_{\tau},
  }
and the known transformation of $\Vev{\O}$. Finally, the $\s_{2n-1}$ functions in~\eqref{pterms} can be expressed in terms of the Eisenstein series $E_2(\tau)$, $E_4(\tau)$ and $E_6(\tau)$,
  \eqsp{
  \s_1 &= \frac{1}{24}[ 1 - E_2(\tau)], \quad \s_3 = \frac{1}{240} [E_4(\tau) - 1], \\
  \s_5 &= \frac{1}{504} [ 1 - E_6(\tau)],
  }
and their modular transformations follow from
  \eq{
  E_2(\tau') &=  (\g \tau + \d)^2 \bigg [ E_2(\tau) + \frac{6}{\pi i} \frac{\g}{\g \tau + \d} \bigg] , \\
  E_{2n}(\tau') &= (\g \tau + \d)^{2n} E_{2n}(\tau), \qquad \mathrm{for}\,\, n \ge 2.
  }
Thus, the modular transformation of  $\Vev{W_0^2}$ is completely determined by the symmetry algebra and the modular invariance of the CFT.

When the contributions to $\Vev{\P_0}$ from the other operators in eq.~\eqref{primary} are taken into account we find
  \eq{
  \Vev{W_0^2}_{\tau', \bar{\tau}'}  =& \Vev{\P_0}_{\tau', \bar{\tau}'} - 2  {\sum_{p=1}^{\infty}}  \Vev{W_{-p}W_{p}}_{\tau',\bar{\tau}'} + \dots \\
  \begin{split}
   \Vev{W_0^2}_{\tau', \bar{\tau}'} =& (\g \tau + \d)^6 \bigg [ \Vev{W_0^2}_{\tau, \bar{\tau}} - \frac{i}{\pi}\frac{\g}{\g \tau + \d} G_{\tau, \bar{\tau}} \\
   & - \frac{2 i}{\pi} \frac{\g}{\g \tau+\d} \Vev{U_0}_{\tau,\bar{\tau}} \bigg ]
   \end{split} \label{secondorder2}
  }
where $G_{\tau,\bar{\tau}}$ is given by
  \eq{
  G_{\tau,\bar{\tau}} = \frac{\xi}{12} \bigg \{&\frac{c}{240} \Big[2 E_4(\tau) + 20 E_2(\tau) + 5 c \Big ] Z(\tau,\bar{\tau},0) \notag \\
    & - \Big [ 2 E_2(\tau) + c \Big] \Vev{L_0}_{\tau, \bar{\tau}} + \Vev{L_0^2}_{\tau, \bar{\tau}} \bigg \}.
  }

Note that eq.~\eqref{secondorder2} holds for any modular invariant CFT with a $\W_{\infty}[\l]$ algebra and hence for $\W_N$ algebras with $N \ge 4$. We can extend this result to CFTs with a $\W_3$ algebra by turning off the dimension-4 current $U(z)$ and removing the contribution of its zero mode from eq.~\eqref{secondorder2}. This works due to the normalization chosen in eq.~\eqref{normalization} and it leads to eqs.~\eqref{secondorder} and~\eqref{G} given in the main text.


\section{\texorpdfstring{$a_n$}{an} and \texorpdfstring{$d_n$}{dn} coefficients}
\label{ap:b}

For completeness, in this section we gather the $a_n$ and $d_n$ coefficients featured in eqs.~\eqref{L2} and~\eqref{barL2}. In finding these expressions we used the following identities for the derivatives of the Eisenstein series
  \eq{
  \p_{\tau} E_2(\tau) &= \frac{\pi i}{6} \big [ E_2(\tau)^2 - E_4(\tau) \big ], \\
  \p_{\tau} E_4(\tau) &= \frac{2\pi i}{3} \big [ E_2(\tau) E_4 (\tau) - E_6(\tau) \big],
  }
along with the explicit values of $E_2(\tau)$ and $E_6(\tau)$ evaluated at $\tau = i$ \phantom{see eq.~\eqref{E4} for the explicit value of $E_4(i)$}
  \eq{
  E_2(i) = \frac{3}{\pi}, \qquad E_6(i) = 0.
  }

The $a_n$ coefficients are given by
  \eq{
  \begin{split}
  a_1 =& \frac{\xi}{10} \Big \{ 315 +15 \pi k \big [66+\pi^2 E_4(i) \big] +\pi^2 k^2 \big[765 \\
  &+ \pi (18+5\pi)E_4(i) \big] + 6 \pi^3 k^3 [30+\pi E_4(i)] \Big \},
  \end{split} \\
  \begin{split}
  a_2 = & -\frac{\pi }{10} \Big \{ 675 +10 \pi k \big [ 216 + \pi^2 E_4(i) \big ] \\
  & + 6 \pi^2 k^2 \big [ 300 + \pi E_4(i) \big ]  + 360 \pi^3 k^3 \Big \}, 
  \end{split} \\
  a_3 =& -18 \pi^3 k^2 , \qquad \qquad \qquad a_4 = {9 \pi^2 k}.
  } 
In contrast, the $d_n$ coefficients are simpler and depend on both central charges,
  \eq{
   d_1 =& 10 - 16 \pi \bar{k} (1+2\pi\bar{k}), \\
   d_2 =&- 32 \pi^3 \Big [ 1+\bar{k}(3+4\pi\bar{k})^2  \Big], \\
   d_5 = & -\frac{4\bar{k}\xi^2}{45} \Big [ 45 + (5 + 6k) \pi^2 E_4(i)  \Big ], \\
   d_3 =& \frac{1+24\pi^2}{8\pi^2}\,d_2, \qquad d_6 = -\frac{\xi}{2\pi} d_2 \\
   d_4 = & - 8 \bar{k} \xi^2, \,\,\, \quad\qquad d_7 =  \frac{1}{16\pi \bar{k}\xi} \, d_2 d_5.
  }

\vspace{-10pt}


\ifprstyle
	\bibliographystyle{apsrev4-1}
\else
	\bibliographystyle{utphys2}
\fi

\bibliography{wcftbound}

\begin{thebibliography}{46}%
\makeatletter
\providecommand \@ifxundefined [1]{%
 \@ifx{#1\undefined}
}%
\providecommand \@ifnum [1]{%
 \ifnum #1\expandafter \@firstoftwo
 \else \expandafter \@secondoftwo
 \fi
}%
\providecommand \@ifx [1]{%
 \ifx #1\expandafter \@firstoftwo
 \else \expandafter \@secondoftwo
 \fi
}%
\providecommand \natexlab [1]{#1}%
\providecommand \enquote  [1]{``#1''}%
\providecommand \bibnamefont  [1]{#1}%
\providecommand \bibfnamefont [1]{#1}%
\providecommand \citenamefont [1]{#1}%
\providecommand \href@noop [0]{\@secondoftwo}%
\providecommand \href [0]{\begingroup \@sanitize@url \@href}%
\providecommand \@href[1]{\@@startlink{#1}\@@href}%
\providecommand \@@href[1]{\endgroup#1\@@endlink}%
\providecommand \@sanitize@url [0]{\catcode `\\12\catcode `\$12\catcode
  `\&12\catcode `\#12\catcode `\^12\catcode `\_12\catcode `\%12\relax}%
\providecommand \@@startlink[1]{}%
\providecommand \@@endlink[0]{}%
\providecommand \url  [0]{\begingroup\@sanitize@url \@url }%
\providecommand \@url [1]{\endgroup\@href {#1}{\urlprefix }}%
\providecommand \urlprefix  [0]{URL }%
\providecommand \Eprint [0]{\href }%
\providecommand \doibase [0]{http://dx.doi.org/}%
\providecommand \selectlanguage [0]{\@gobble}%
\providecommand \bibinfo  [0]{\@secondoftwo}%
\providecommand \bibfield  [0]{\@secondoftwo}%
\providecommand \translation [1]{[#1]}%
\providecommand \BibitemOpen [0]{}%
\providecommand \bibitemStop [0]{}%
\providecommand \bibitemNoStop [0]{.\EOS\space}%
\providecommand \EOS [0]{\spacefactor3000\relax}%
\providecommand \BibitemShut  [1]{\csname bibitem#1\endcsname}%
\let\auto@bib@innerbib\@empty
\bibitem [{\citenamefont {Belavin}\ \emph {et~al.}(1984)\citenamefont
  {Belavin}, \citenamefont {Polyakov},\ and\ \citenamefont
  {Zamolodchikov}}]{Belavin:1984vu}%
  \BibitemOpen
  \bibfield  {author} {\bibinfo {author} {\bibfnamefont {A.~A.}\ \bibnamefont
  {Belavin}}, \bibinfo {author} {\bibfnamefont {A.~M.}\ \bibnamefont
  {Polyakov}}, \ and\ \bibinfo {author} {\bibfnamefont {A.~B.}\ \bibnamefont
  {Zamolodchikov}},\ }\href {\doibase 10.1016/0550-3213(84)90052-X} {\bibfield
  {journal} {\bibinfo  {journal} {Nucl. Phys.}\ }\textbf {\bibinfo {volume}
  {B241}},\ \bibinfo {pages} {333} (\bibinfo {year} {1984})}\BibitemShut
  {NoStop}%
\bibitem [{\citenamefont {Cardy}(1991)}]{Cardy:1991kr}%
  \BibitemOpen
  \bibfield  {author} {\bibinfo {author} {\bibfnamefont {J.~L.}\ \bibnamefont
  {Cardy}},\ }\href {\doibase 10.1016/0550-3213(91)90024-R} {\bibfield
  {journal} {\bibinfo  {journal} {Nucl. Phys.}\ }\textbf {\bibinfo {volume}
  {B366}},\ \bibinfo {pages} {403} (\bibinfo {year} {1991})}\BibitemShut
  {NoStop}%
\bibitem [{\citenamefont {Belin}\ \emph {et~al.}(2017)\citenamefont {Belin},
  \citenamefont {de~Boer}, \citenamefont {Kruthoff}, \citenamefont {Michel},
  \citenamefont {Shaghoulian},\ and\ \citenamefont {Shyani}}]{Belin:2016yll}%
  \BibitemOpen
  \bibfield  {author} {\bibinfo {author} {\bibfnamefont {A.}~\bibnamefont
  {Belin}}, \bibinfo {author} {\bibfnamefont {J.}~\bibnamefont {de~Boer}},
  \bibinfo {author} {\bibfnamefont {J.}~\bibnamefont {Kruthoff}}, \bibinfo
  {author} {\bibfnamefont {B.}~\bibnamefont {Michel}}, \bibinfo {author}
  {\bibfnamefont {E.}~\bibnamefont {Shaghoulian}}, \ and\ \bibinfo {author}
  {\bibfnamefont {M.}~\bibnamefont {Shyani}},\ }\href {\doibase
  10.1007/JHEP03(2017)067} {\bibfield  {journal} {\bibinfo  {journal} {JHEP}\
  }\textbf {\bibinfo {volume} {03}},\ \bibinfo {pages} {067} (\bibinfo {year}
  {2017})},\ \Eprint {http://arxiv.org/abs/1610.06186} {arXiv:1610.06186
  [hep-th]} \BibitemShut {NoStop}%
\bibitem [{\citenamefont {Shaghoulian}(2016)}]{Shaghoulian:2016gol}%
  \BibitemOpen
  \bibfield  {author} {\bibinfo {author} {\bibfnamefont {E.}~\bibnamefont
  {Shaghoulian}},\ }\href@noop {} {\  (\bibinfo {year} {2016})},\ \Eprint
  {http://arxiv.org/abs/1612.05257} {arXiv:1612.05257 [hep-th]} \BibitemShut
  {NoStop}%
\bibitem [{\citenamefont {Rattazzi}\ \emph {et~al.}(2008)\citenamefont
  {Rattazzi}, \citenamefont {Rychkov}, \citenamefont {Tonni},\ and\
  \citenamefont {Vichi}}]{Rattazzi:2008pe}%
  \BibitemOpen
  \bibfield  {author} {\bibinfo {author} {\bibfnamefont {R.}~\bibnamefont
  {Rattazzi}}, \bibinfo {author} {\bibfnamefont {V.~S.}\ \bibnamefont
  {Rychkov}}, \bibinfo {author} {\bibfnamefont {E.}~\bibnamefont {Tonni}}, \
  and\ \bibinfo {author} {\bibfnamefont {A.}~\bibnamefont {Vichi}},\ }\href
  {\doibase 10.1088/1126-6708/2008/12/031} {\bibfield  {journal} {\bibinfo
  {journal} {JHEP}\ }\textbf {\bibinfo {volume} {12}},\ \bibinfo {pages} {031}
  (\bibinfo {year} {2008})},\ \Eprint {http://arxiv.org/abs/0807.0004}
  {arXiv:0807.0004 [hep-th]} \BibitemShut {NoStop}%
\bibitem [{\citenamefont {Poland}\ \emph {et~al.}(2012)\citenamefont {Poland},
  \citenamefont {Simmons-Duffin},\ and\ \citenamefont {Vichi}}]{Poland:2011ey}%
  \BibitemOpen
  \bibfield  {author} {\bibinfo {author} {\bibfnamefont {D.}~\bibnamefont
  {Poland}}, \bibinfo {author} {\bibfnamefont {D.}~\bibnamefont
  {Simmons-Duffin}}, \ and\ \bibinfo {author} {\bibfnamefont {A.}~\bibnamefont
  {Vichi}},\ }\href {\doibase 10.1007/JHEP05(2012)110} {\bibfield  {journal}
  {\bibinfo  {journal} {JHEP}\ }\textbf {\bibinfo {volume} {05}},\ \bibinfo
  {pages} {110} (\bibinfo {year} {2012})},\ \Eprint
  {http://arxiv.org/abs/1109.5176} {arXiv:1109.5176 [hep-th]} \BibitemShut
  {NoStop}%
\bibitem [{\citenamefont {El-Showk}\ \emph {et~al.}(2012)\citenamefont
  {El-Showk}, \citenamefont {Paulos}, \citenamefont {Poland}, \citenamefont
  {Rychkov}, \citenamefont {Simmons-Duffin},\ and\ \citenamefont
  {Vichi}}]{ElShowk:2012ht}%
  \BibitemOpen
  \bibfield  {author} {\bibinfo {author} {\bibfnamefont {S.}~\bibnamefont
  {El-Showk}}, \bibinfo {author} {\bibfnamefont {M.~F.}\ \bibnamefont
  {Paulos}}, \bibinfo {author} {\bibfnamefont {D.}~\bibnamefont {Poland}},
  \bibinfo {author} {\bibfnamefont {S.}~\bibnamefont {Rychkov}}, \bibinfo
  {author} {\bibfnamefont {D.}~\bibnamefont {Simmons-Duffin}}, \ and\ \bibinfo
  {author} {\bibfnamefont {A.}~\bibnamefont {Vichi}},\ }\href {\doibase
  10.1103/PhysRevD.86.025022} {\bibfield  {journal} {\bibinfo  {journal} {Phys.
  Rev.}\ }\textbf {\bibinfo {volume} {D86}},\ \bibinfo {pages} {025022}
  (\bibinfo {year} {2012})},\ \Eprint {http://arxiv.org/abs/1203.6064}
  {arXiv:1203.6064 [hep-th]} \BibitemShut {NoStop}%
\bibitem [{\citenamefont {Cardy}(1986)}]{Cardy:1986ie}%
  \BibitemOpen
  \bibfield  {author} {\bibinfo {author} {\bibfnamefont {J.~L.}\ \bibnamefont
  {Cardy}},\ }\href {\doibase 10.1016/0550-3213(86)90552-3} {\bibfield
  {journal} {\bibinfo  {journal} {Nucl. Phys.}\ }\textbf {\bibinfo {volume}
  {B270}},\ \bibinfo {pages} {186} (\bibinfo {year} {1986})}\BibitemShut
  {NoStop}%
\bibitem [{\citenamefont {Hellerman}(2011)}]{Hellerman:2009bu}%
  \BibitemOpen
  \bibfield  {author} {\bibinfo {author} {\bibfnamefont {S.}~\bibnamefont
  {Hellerman}},\ }\href {\doibase 10.1007/JHEP08(2011)130} {\bibfield
  {journal} {\bibinfo  {journal} {JHEP}\ }\textbf {\bibinfo {volume} {08}},\
  \bibinfo {pages} {130} (\bibinfo {year} {2011})},\ \Eprint
  {http://arxiv.org/abs/0902.2790} {arXiv:0902.2790 [hep-th]} \BibitemShut
  {NoStop}%
\bibitem [{\citenamefont {Hellerman}\ and\ \citenamefont
  {Schmidt-Colinet}(2011)}]{Hellerman:2010qd}%
  \BibitemOpen
  \bibfield  {author} {\bibinfo {author} {\bibfnamefont {S.}~\bibnamefont
  {Hellerman}}\ and\ \bibinfo {author} {\bibfnamefont {C.}~\bibnamefont
  {Schmidt-Colinet}},\ }\href {\doibase 10.1007/JHEP08(2011)127} {\bibfield
  {journal} {\bibinfo  {journal} {JHEP}\ }\textbf {\bibinfo {volume} {08}},\
  \bibinfo {pages} {127} (\bibinfo {year} {2011})},\ \Eprint
  {http://arxiv.org/abs/1007.0756} {arXiv:1007.0756 [hep-th]} \BibitemShut
  {NoStop}%
\bibitem [{\citenamefont {Keller}\ and\ \citenamefont
  {Ooguri}(2013)}]{Keller:2012mr}%
  \BibitemOpen
  \bibfield  {author} {\bibinfo {author} {\bibfnamefont {C.~A.}\ \bibnamefont
  {Keller}}\ and\ \bibinfo {author} {\bibfnamefont {H.}~\bibnamefont
  {Ooguri}},\ }\href {\doibase 10.1007/s00220-013-1797-8} {\bibfield  {journal}
  {\bibinfo  {journal} {Commun. Math. Phys.}\ }\textbf {\bibinfo {volume}
  {324}},\ \bibinfo {pages} {107} (\bibinfo {year} {2013})},\ \Eprint
  {http://arxiv.org/abs/1209.4649} {arXiv:1209.4649 [hep-th]} \BibitemShut
  {NoStop}%
\bibitem [{\citenamefont {Friedan}\ and\ \citenamefont
  {Keller}(2013)}]{Friedan:2013cba}%
  \BibitemOpen
  \bibfield  {author} {\bibinfo {author} {\bibfnamefont {D.}~\bibnamefont
  {Friedan}}\ and\ \bibinfo {author} {\bibfnamefont {C.~A.}\ \bibnamefont
  {Keller}},\ }\href {\doibase 10.1007/JHEP10(2013)180} {\bibfield  {journal}
  {\bibinfo  {journal} {JHEP}\ }\textbf {\bibinfo {volume} {10}},\ \bibinfo
  {pages} {180} (\bibinfo {year} {2013})},\ \Eprint
  {http://arxiv.org/abs/1307.6562} {arXiv:1307.6562 [hep-th]} \BibitemShut
  {NoStop}%
\bibitem [{\citenamefont {Qualls}\ and\ \citenamefont
  {Shapere}(2014)}]{Qualls:2013eha}%
  \BibitemOpen
  \bibfield  {author} {\bibinfo {author} {\bibfnamefont {J.~D.}\ \bibnamefont
  {Qualls}}\ and\ \bibinfo {author} {\bibfnamefont {A.~D.}\ \bibnamefont
  {Shapere}},\ }\href {\doibase 10.1007/JHEP05(2014)091} {\bibfield  {journal}
  {\bibinfo  {journal} {JHEP}\ }\textbf {\bibinfo {volume} {05}},\ \bibinfo
  {pages} {091} (\bibinfo {year} {2014})},\ \Eprint
  {http://arxiv.org/abs/1312.0038} {arXiv:1312.0038 [hep-th]} \BibitemShut
  {NoStop}%
\bibitem [{\citenamefont {Qualls}(2015{\natexlab{a}})}]{Qualls:2014oea}%
  \BibitemOpen
  \bibfield  {author} {\bibinfo {author} {\bibfnamefont {J.~D.}\ \bibnamefont
  {Qualls}},\ }\href {\doibase 10.1007/JHEP12(2015)001} {\bibfield  {journal}
  {\bibinfo  {journal} {JHEP}\ }\textbf {\bibinfo {volume} {12}},\ \bibinfo
  {pages} {001} (\bibinfo {year} {2015}{\natexlab{a}})},\ \Eprint
  {http://arxiv.org/abs/1412.0383} {arXiv:1412.0383 [hep-th]} \BibitemShut
  {NoStop}%
\bibitem [{\citenamefont {Qualls}(2015{\natexlab{b}})}]{Qualls:2015bta}%
  \BibitemOpen
  \bibfield  {author} {\bibinfo {author} {\bibfnamefont {J.~D.}\ \bibnamefont
  {Qualls}},\ }\href@noop {} {\  (\bibinfo {year} {2015}{\natexlab{b}})},\
  \Eprint {http://arxiv.org/abs/1508.00548} {arXiv:1508.00548 [hep-th]}
  \BibitemShut {NoStop}%
\bibitem [{\citenamefont {Kim}\ \emph {et~al.}(2016)\citenamefont {Kim},
  \citenamefont {Kravchuk},\ and\ \citenamefont {Ooguri}}]{Kim:2015oca}%
  \BibitemOpen
  \bibfield  {author} {\bibinfo {author} {\bibfnamefont {H.}~\bibnamefont
  {Kim}}, \bibinfo {author} {\bibfnamefont {P.}~\bibnamefont {Kravchuk}}, \
  and\ \bibinfo {author} {\bibfnamefont {H.}~\bibnamefont {Ooguri}},\ }\href
  {\doibase 10.1007/JHEP04(2016)184} {\bibfield  {journal} {\bibinfo  {journal}
  {JHEP}\ }\textbf {\bibinfo {volume} {04}},\ \bibinfo {pages} {184} (\bibinfo
  {year} {2016})},\ \Eprint {http://arxiv.org/abs/1510.08772} {arXiv:1510.08772
  [hep-th]} \BibitemShut {NoStop}%
\bibitem [{\citenamefont {Benjamin}\ \emph {et~al.}(2016)\citenamefont
  {Benjamin}, \citenamefont {Dyer}, \citenamefont {Fitzpatrick},\ and\
  \citenamefont {Kachru}}]{Benjamin:2016fhe}%
  \BibitemOpen
  \bibfield  {author} {\bibinfo {author} {\bibfnamefont {N.}~\bibnamefont
  {Benjamin}}, \bibinfo {author} {\bibfnamefont {E.}~\bibnamefont {Dyer}},
  \bibinfo {author} {\bibfnamefont {A.~L.}\ \bibnamefont {Fitzpatrick}}, \ and\
  \bibinfo {author} {\bibfnamefont {S.}~\bibnamefont {Kachru}},\ }\href
  {\doibase 10.1007/JHEP08(2016)041} {\bibfield  {journal} {\bibinfo  {journal}
  {JHEP}\ }\textbf {\bibinfo {volume} {08}},\ \bibinfo {pages} {041} (\bibinfo
  {year} {2016})},\ \Eprint {http://arxiv.org/abs/1603.09745} {arXiv:1603.09745
  [hep-th]} \BibitemShut {NoStop}%
\bibitem [{\citenamefont {Collier}\ \emph {et~al.}(2016)\citenamefont
  {Collier}, \citenamefont {Lin},\ and\ \citenamefont {Yin}}]{Collier:2016cls}%
  \BibitemOpen
  \bibfield  {author} {\bibinfo {author} {\bibfnamefont {S.}~\bibnamefont
  {Collier}}, \bibinfo {author} {\bibfnamefont {Y.-H.}\ \bibnamefont {Lin}}, \
  and\ \bibinfo {author} {\bibfnamefont {X.}~\bibnamefont {Yin}},\ }\href@noop
  {} {\  (\bibinfo {year} {2016})},\ \Eprint {http://arxiv.org/abs/1608.06241}
  {arXiv:1608.06241 [hep-th]} \BibitemShut {NoStop}%
\bibitem [{\citenamefont {Campoleoni}\ \emph {et~al.}(2010)\citenamefont
  {Campoleoni}, \citenamefont {Fredenhagen}, \citenamefont {Pfenninger},\ and\
  \citenamefont {Theisen}}]{Campoleoni:2010zq}%
  \BibitemOpen
  \bibfield  {author} {\bibinfo {author} {\bibfnamefont {A.}~\bibnamefont
  {Campoleoni}}, \bibinfo {author} {\bibfnamefont {S.}~\bibnamefont
  {Fredenhagen}}, \bibinfo {author} {\bibfnamefont {S.}~\bibnamefont
  {Pfenninger}}, \ and\ \bibinfo {author} {\bibfnamefont {S.}~\bibnamefont
  {Theisen}},\ }\href {\doibase 10.1007/JHEP11(2010)007} {\bibfield  {journal}
  {\bibinfo  {journal} {JHEP}\ }\textbf {\bibinfo {volume} {11}},\ \bibinfo
  {pages} {007} (\bibinfo {year} {2010})},\ \Eprint
  {http://arxiv.org/abs/1008.4744} {arXiv:1008.4744 [hep-th]} \BibitemShut
  {NoStop}%
\bibitem [{\citenamefont {Maldacena}(1999)}]{Maldacena:1997re}%
  \BibitemOpen
  \bibfield  {author} {\bibinfo {author} {\bibfnamefont {J.~M.}\ \bibnamefont
  {Maldacena}},\ }\href {\doibase 10.1023/A:1026654312961} {\bibfield
  {journal} {\bibinfo  {journal} {Int.J.Theor.Phys.}\ }\textbf {\bibinfo
  {volume} {38}},\ \bibinfo {pages} {1113} (\bibinfo {year} {1999})},\ \Eprint
  {http://arxiv.org/abs/hep-th/9711200} {arXiv:hep-th/9711200 [hep-th]}
  \BibitemShut {NoStop}%
\bibitem [{\citenamefont {Gubser}\ \emph {et~al.}(1998)\citenamefont {Gubser},
  \citenamefont {Klebanov},\ and\ \citenamefont {Polyakov}}]{Gubser:1998bc}%
  \BibitemOpen
  \bibfield  {author} {\bibinfo {author} {\bibfnamefont {S.}~\bibnamefont
  {Gubser}}, \bibinfo {author} {\bibfnamefont {I.~R.}\ \bibnamefont
  {Klebanov}}, \ and\ \bibinfo {author} {\bibfnamefont {A.~M.}\ \bibnamefont
  {Polyakov}},\ }\href {\doibase 10.1016/S0370-2693(98)00377-3} {\bibfield
  {journal} {\bibinfo  {journal} {Phys.Lett.}\ }\textbf {\bibinfo {volume}
  {B428}},\ \bibinfo {pages} {105} (\bibinfo {year} {1998})},\ \Eprint
  {http://arxiv.org/abs/hep-th/9802109} {arXiv:hep-th/9802109 [hep-th]}
  \BibitemShut {NoStop}%
\bibitem [{\citenamefont {Witten}(1998)}]{Witten:1998qj}%
  \BibitemOpen
  \bibfield  {author} {\bibinfo {author} {\bibfnamefont {E.}~\bibnamefont
  {Witten}},\ }\href@noop {} {\bibfield  {journal} {\bibinfo  {journal}
  {Adv.Theor.Math.Phys.}\ }\textbf {\bibinfo {volume} {2}},\ \bibinfo {pages}
  {253} (\bibinfo {year} {1998})},\ \Eprint
  {http://arxiv.org/abs/hep-th/9802150} {arXiv:hep-th/9802150 [hep-th]}
  \BibitemShut {NoStop}%
\bibitem [{\citenamefont {Henneaux}\ and\ \citenamefont
  {Rey}(2010)}]{Henneaux:2010xg}%
  \BibitemOpen
  \bibfield  {author} {\bibinfo {author} {\bibfnamefont {M.}~\bibnamefont
  {Henneaux}}\ and\ \bibinfo {author} {\bibfnamefont {S.-J.}\ \bibnamefont
  {Rey}},\ }\href {\doibase 10.1007/JHEP12(2010)007} {\bibfield  {journal}
  {\bibinfo  {journal} {JHEP}\ }\textbf {\bibinfo {volume} {12}},\ \bibinfo
  {pages} {007} (\bibinfo {year} {2010})},\ \Eprint
  {http://arxiv.org/abs/1008.4579} {arXiv:1008.4579 [hep-th]} \BibitemShut
  {NoStop}%
\bibitem [{\citenamefont {Campoleoni}\ \emph {et~al.}(2011)\citenamefont
  {Campoleoni}, \citenamefont {Fredenhagen},\ and\ \citenamefont
  {Pfenninger}}]{Campoleoni:2011hg}%
  \BibitemOpen
  \bibfield  {author} {\bibinfo {author} {\bibfnamefont {A.}~\bibnamefont
  {Campoleoni}}, \bibinfo {author} {\bibfnamefont {S.}~\bibnamefont
  {Fredenhagen}}, \ and\ \bibinfo {author} {\bibfnamefont {S.}~\bibnamefont
  {Pfenninger}},\ }\href {\doibase 10.1007/JHEP09(2011)113} {\bibfield
  {journal} {\bibinfo  {journal} {JHEP}\ }\textbf {\bibinfo {volume} {09}},\
  \bibinfo {pages} {113} (\bibinfo {year} {2011})},\ \Eprint
  {http://arxiv.org/abs/1107.0290} {arXiv:1107.0290 [hep-th]} \BibitemShut
  {NoStop}%
\bibitem [{\citenamefont {Gaberdiel}\ and\ \citenamefont
  {Gopakumar}(2011)}]{Gaberdiel:2010pz}%
  \BibitemOpen
  \bibfield  {author} {\bibinfo {author} {\bibfnamefont {M.~R.}\ \bibnamefont
  {Gaberdiel}}\ and\ \bibinfo {author} {\bibfnamefont {R.}~\bibnamefont
  {Gopakumar}},\ }\href {\doibase 10.1103/PhysRevD.83.066007} {\bibfield
  {journal} {\bibinfo  {journal} {Phys. Rev.}\ }\textbf {\bibinfo {volume}
  {D83}},\ \bibinfo {pages} {066007} (\bibinfo {year} {2011})},\ \Eprint
  {http://arxiv.org/abs/1011.2986} {arXiv:1011.2986 [hep-th]} \BibitemShut
  {NoStop}%
\bibitem [{\citenamefont {Gaberdiel}\ and\ \citenamefont
  {Gopakumar}(2013)}]{Gaberdiel:2012uj}%
  \BibitemOpen
  \bibfield  {author} {\bibinfo {author} {\bibfnamefont {M.~R.}\ \bibnamefont
  {Gaberdiel}}\ and\ \bibinfo {author} {\bibfnamefont {R.}~\bibnamefont
  {Gopakumar}},\ }\href {\doibase 10.1088/1751-8113/46/21/214002} {\bibfield
  {journal} {\bibinfo  {journal} {J. Phys.}\ }\textbf {\bibinfo {volume}
  {A46}},\ \bibinfo {pages} {214002} (\bibinfo {year} {2013})},\ \Eprint
  {http://arxiv.org/abs/1207.6697} {arXiv:1207.6697 [hep-th]} \BibitemShut
  {NoStop}%
\bibitem [{\citenamefont {Witten}(2007)}]{Witten:2007kt}%
  \BibitemOpen
  \bibfield  {author} {\bibinfo {author} {\bibfnamefont {E.}~\bibnamefont
  {Witten}},\ }\href@noop {} {\  (\bibinfo {year} {2007})},\ \Eprint
  {http://arxiv.org/abs/0706.3359} {arXiv:0706.3359 [hep-th]} \BibitemShut
  {NoStop}%
\bibitem [{\citenamefont {Gaberdiel}\ \emph {et~al.}(2008)\citenamefont
  {Gaberdiel}, \citenamefont {Gukov}, \citenamefont {Keller}, \citenamefont
  {Moore},\ and\ \citenamefont {Ooguri}}]{Gaberdiel:2008xb}%
  \BibitemOpen
  \bibfield  {author} {\bibinfo {author} {\bibfnamefont {M.~R.}\ \bibnamefont
  {Gaberdiel}}, \bibinfo {author} {\bibfnamefont {S.}~\bibnamefont {Gukov}},
  \bibinfo {author} {\bibfnamefont {C.~A.}\ \bibnamefont {Keller}}, \bibinfo
  {author} {\bibfnamefont {G.~W.}\ \bibnamefont {Moore}}, \ and\ \bibinfo
  {author} {\bibfnamefont {H.}~\bibnamefont {Ooguri}},\ }\href {\doibase
  10.4310/CNTP.2008.v2.n4.a3} {\bibfield  {journal} {\bibinfo  {journal}
  {Commun. Num. Theor. Phys.}\ }\textbf {\bibinfo {volume} {2}},\ \bibinfo
  {pages} {743} (\bibinfo {year} {2008})},\ \Eprint
  {http://arxiv.org/abs/0805.4216} {arXiv:0805.4216 [hep-th]} \BibitemShut
  {NoStop}%
\bibitem [{\citenamefont {Gutperle}\ and\ \citenamefont
  {Kraus}(2011)}]{Gutperle:2011kf}%
  \BibitemOpen
  \bibfield  {author} {\bibinfo {author} {\bibfnamefont {M.}~\bibnamefont
  {Gutperle}}\ and\ \bibinfo {author} {\bibfnamefont {P.}~\bibnamefont
  {Kraus}},\ }\href {\doibase 10.1007/JHEP05(2011)022} {\bibfield  {journal}
  {\bibinfo  {journal} {JHEP}\ }\textbf {\bibinfo {volume} {05}},\ \bibinfo
  {pages} {022} (\bibinfo {year} {2011})},\ \Eprint
  {http://arxiv.org/abs/1103.4304} {arXiv:1103.4304 [hep-th]} \BibitemShut
  {NoStop}%
\bibitem [{\citenamefont {Ammon}\ \emph {et~al.}(2013)\citenamefont {Ammon},
  \citenamefont {Gutperle}, \citenamefont {Kraus},\ and\ \citenamefont
  {Perlmutter}}]{Ammon:2012wc}%
  \BibitemOpen
  \bibfield  {author} {\bibinfo {author} {\bibfnamefont {M.}~\bibnamefont
  {Ammon}}, \bibinfo {author} {\bibfnamefont {M.}~\bibnamefont {Gutperle}},
  \bibinfo {author} {\bibfnamefont {P.}~\bibnamefont {Kraus}}, \ and\ \bibinfo
  {author} {\bibfnamefont {E.}~\bibnamefont {Perlmutter}},\ }\href {\doibase
  10.1088/1751-8113/46/21/214001} {\bibfield  {journal} {\bibinfo  {journal}
  {J. Phys.}\ }\textbf {\bibinfo {volume} {A46}},\ \bibinfo {pages} {214001}
  (\bibinfo {year} {2013})},\ \Eprint {http://arxiv.org/abs/1208.5182}
  {arXiv:1208.5182 [hep-th]} \BibitemShut {NoStop}%
\bibitem [{\citenamefont {Bunster}\ \emph {et~al.}(2014)\citenamefont
  {Bunster}, \citenamefont {Henneaux}, \citenamefont {Perez}, \citenamefont
  {Tempo},\ and\ \citenamefont {Troncoso}}]{Bunster:2014mua}%
  \BibitemOpen
  \bibfield  {author} {\bibinfo {author} {\bibfnamefont {C.}~\bibnamefont
  {Bunster}}, \bibinfo {author} {\bibfnamefont {M.}~\bibnamefont {Henneaux}},
  \bibinfo {author} {\bibfnamefont {A.}~\bibnamefont {Perez}}, \bibinfo
  {author} {\bibfnamefont {D.}~\bibnamefont {Tempo}}, \ and\ \bibinfo {author}
  {\bibfnamefont {R.}~\bibnamefont {Troncoso}},\ }\href {\doibase
  10.1007/JHEP05(2014)031} {\bibfield  {journal} {\bibinfo  {journal} {JHEP}\
  }\textbf {\bibinfo {volume} {05}},\ \bibinfo {pages} {031} (\bibinfo {year}
  {2014})},\ \Eprint {http://arxiv.org/abs/1404.3305} {arXiv:1404.3305
  [hep-th]} \BibitemShut {NoStop}%
\bibitem [{\citenamefont {Ba\~nados}\ \emph {et~al.}(2016)\citenamefont
  {Ba\~nados}, \citenamefont {Castro}, \citenamefont {Faraggi},\ and\
  \citenamefont {Jottar}}]{Banados:2015tft}%
  \BibitemOpen
  \bibfield  {author} {\bibinfo {author} {\bibfnamefont {M.}~\bibnamefont
  {Ba\~nados}}, \bibinfo {author} {\bibfnamefont {A.}~\bibnamefont {Castro}},
  \bibinfo {author} {\bibfnamefont {A.}~\bibnamefont {Faraggi}}, \ and\
  \bibinfo {author} {\bibfnamefont {J.~I.}\ \bibnamefont {Jottar}},\ }\href
  {\doibase 10.1007/JHEP04(2016)077} {\bibfield  {journal} {\bibinfo  {journal}
  {JHEP}\ }\textbf {\bibinfo {volume} {04}},\ \bibinfo {pages} {077} (\bibinfo
  {year} {2016})},\ \Eprint {http://arxiv.org/abs/1512.00073} {arXiv:1512.00073
  [hep-th]} \BibitemShut {NoStop}%
\bibitem [{\citenamefont {Arkani-Hamed}\ \emph {et~al.}(2007)\citenamefont
  {Arkani-Hamed}, \citenamefont {Motl}, \citenamefont {Nicolis},\ and\
  \citenamefont {Vafa}}]{ArkaniHamed:2006dz}%
  \BibitemOpen
  \bibfield  {author} {\bibinfo {author} {\bibfnamefont {N.}~\bibnamefont
  {Arkani-Hamed}}, \bibinfo {author} {\bibfnamefont {L.}~\bibnamefont {Motl}},
  \bibinfo {author} {\bibfnamefont {A.}~\bibnamefont {Nicolis}}, \ and\
  \bibinfo {author} {\bibfnamefont {C.}~\bibnamefont {Vafa}},\ }\href {\doibase
  10.1088/1126-6708/2007/06/060} {\bibfield  {journal} {\bibinfo  {journal}
  {JHEP}\ }\textbf {\bibinfo {volume} {06}},\ \bibinfo {pages} {060} (\bibinfo
  {year} {2007})},\ \Eprint {http://arxiv.org/abs/hep-th/0601001}
  {arXiv:hep-th/0601001 [hep-th]} \BibitemShut {NoStop}%
\bibitem [{\citenamefont {Nakayama}\ and\ \citenamefont
  {Nomura}(2015)}]{Nakayama:2015hga}%
  \BibitemOpen
  \bibfield  {author} {\bibinfo {author} {\bibfnamefont {Y.}~\bibnamefont
  {Nakayama}}\ and\ \bibinfo {author} {\bibfnamefont {Y.}~\bibnamefont
  {Nomura}},\ }\href {\doibase 10.1103/PhysRevD.92.126006} {\bibfield
  {journal} {\bibinfo  {journal} {Phys. Rev.}\ }\textbf {\bibinfo {volume}
  {D92}},\ \bibinfo {pages} {126006} (\bibinfo {year} {2015})},\ \Eprint
  {http://arxiv.org/abs/1509.01647} {arXiv:1509.01647 [hep-th]} \BibitemShut
  {NoStop}%
\bibitem [{\citenamefont {Montero}\ \emph {et~al.}(2016)\citenamefont
  {Montero}, \citenamefont {Shiu},\ and\ \citenamefont
  {Soler}}]{Montero:2016tif}%
  \BibitemOpen
  \bibfield  {author} {\bibinfo {author} {\bibfnamefont {M.}~\bibnamefont
  {Montero}}, \bibinfo {author} {\bibfnamefont {G.}~\bibnamefont {Shiu}}, \
  and\ \bibinfo {author} {\bibfnamefont {P.}~\bibnamefont {Soler}},\ }\href
  {\doibase 10.1007/JHEP10(2016)159} {\bibfield  {journal} {\bibinfo  {journal}
  {JHEP}\ }\textbf {\bibinfo {volume} {10}},\ \bibinfo {pages} {159} (\bibinfo
  {year} {2016})},\ \Eprint {http://arxiv.org/abs/1606.08438} {arXiv:1606.08438
  [hep-th]} \BibitemShut {NoStop}%
\bibitem [{\citenamefont {Perlmutter}(2016)}]{Perlmutter:2016pkf}%
  \BibitemOpen
  \bibfield  {author} {\bibinfo {author} {\bibfnamefont {E.}~\bibnamefont
  {Perlmutter}},\ }\href {\doibase 10.1007/JHEP10(2016)069} {\bibfield
  {journal} {\bibinfo  {journal} {JHEP}\ }\textbf {\bibinfo {volume} {10}},\
  \bibinfo {pages} {069} (\bibinfo {year} {2016})},\ \Eprint
  {http://arxiv.org/abs/1602.08272} {arXiv:1602.08272 [hep-th]} \BibitemShut
  {NoStop}%
\bibitem [{\citenamefont {Zamolodchikov}(1985)}]{Zamolodchikov:1985wn}%
  \BibitemOpen
  \bibfield  {author} {\bibinfo {author} {\bibfnamefont {A.~B.}\ \bibnamefont
  {Zamolodchikov}},\ }\href {\doibase 10.1007/BF01036128} {\bibfield  {journal}
  {\bibinfo  {journal} {Theor. Math. Phys.}\ }\textbf {\bibinfo {volume}
  {65}},\ \bibinfo {pages} {1205} (\bibinfo {year} {1985})},\ \bibinfo {note}
  {[Teor. Mat. Fiz.65,347(1985)]}\BibitemShut {NoStop}%
\bibitem [{\citenamefont {Gaberdiel}\ \emph {et~al.}(2012)\citenamefont
  {Gaberdiel}, \citenamefont {Hartman},\ and\ \citenamefont
  {Jin}}]{Gaberdiel:2012yb}%
  \BibitemOpen
  \bibfield  {author} {\bibinfo {author} {\bibfnamefont {M.~R.}\ \bibnamefont
  {Gaberdiel}}, \bibinfo {author} {\bibfnamefont {T.}~\bibnamefont {Hartman}},
  \ and\ \bibinfo {author} {\bibfnamefont {K.}~\bibnamefont {Jin}},\ }\href
  {\doibase 10.1007/JHEP04(2012)103} {\bibfield  {journal} {\bibinfo  {journal}
  {JHEP}\ }\textbf {\bibinfo {volume} {04}},\ \bibinfo {pages} {103} (\bibinfo
  {year} {2012})},\ \Eprint {http://arxiv.org/abs/1203.0015} {arXiv:1203.0015
  [hep-th]} \BibitemShut {NoStop}%
\bibitem [{\citenamefont {Iles}\ and\ \citenamefont
  {Watts}(2016)}]{Iles:2014gra}%
  \BibitemOpen
  \bibfield  {author} {\bibinfo {author} {\bibfnamefont {N.~J.}\ \bibnamefont
  {Iles}}\ and\ \bibinfo {author} {\bibfnamefont {G.~M.~T.}\ \bibnamefont
  {Watts}},\ }\href {\doibase 10.1007/JHEP01(2016)089} {\bibfield  {journal}
  {\bibinfo  {journal} {JHEP}\ }\textbf {\bibinfo {volume} {01}},\ \bibinfo
  {pages} {089} (\bibinfo {year} {2016})},\ \Eprint
  {http://arxiv.org/abs/1411.4039} {arXiv:1411.4039 [hep-th]} \BibitemShut
  {NoStop}%
\bibitem [{\citenamefont {Li}\ \emph {et~al.}(2013)\citenamefont {Li},
  \citenamefont {Lin},\ and\ \citenamefont {Wang}}]{Li:2013rsa}%
  \BibitemOpen
  \bibfield  {author} {\bibinfo {author} {\bibfnamefont {W.}~\bibnamefont
  {Li}}, \bibinfo {author} {\bibfnamefont {F.-L.}\ \bibnamefont {Lin}}, \ and\
  \bibinfo {author} {\bibfnamefont {C.-W.}\ \bibnamefont {Wang}},\ }\href
  {\doibase 10.1007/JHEP12(2013)094} {\bibfield  {journal} {\bibinfo  {journal}
  {JHEP}\ }\textbf {\bibinfo {volume} {12}},\ \bibinfo {pages} {094} (\bibinfo
  {year} {2013})},\ \Eprint {http://arxiv.org/abs/1308.2959} {arXiv:1308.2959
  [hep-th]} \BibitemShut {NoStop}%
\bibitem [{\citenamefont {Kaneko}\ and\ \citenamefont
  {Zagier}(1995)}]{Kaneko:1995abc}%
  \BibitemOpen
  \bibfield  {author} {\bibinfo {author} {\bibfnamefont {M.}~\bibnamefont
  {Kaneko}}\ and\ \bibinfo {author} {\bibfnamefont {D.}~\bibnamefont
  {Zagier}},\ }in\ \href@noop {} {\emph {\bibinfo {booktitle} {{The Moduli
  Space of Curves}}}}\ (\bibinfo  {publisher} {Birkh\"auser Boston},\ \bibinfo
  {year} {1995})\BibitemShut {NoStop}%
\bibitem [{\citenamefont {Iles}\ and\ \citenamefont
  {Watts}(2014)}]{Iles:2013jha}%
  \BibitemOpen
  \bibfield  {author} {\bibinfo {author} {\bibfnamefont {N.~J.}\ \bibnamefont
  {Iles}}\ and\ \bibinfo {author} {\bibfnamefont {G.~M.~T.}\ \bibnamefont
  {Watts}},\ }\href {\doibase 10.1007/JHEP02(2014)009} {\bibfield  {journal}
  {\bibinfo  {journal} {JHEP}\ }\textbf {\bibinfo {volume} {02}},\ \bibinfo
  {pages} {009} (\bibinfo {year} {2014})},\ \Eprint
  {http://arxiv.org/abs/1307.3771} {arXiv:1307.3771 [hep-th]} \BibitemShut
  {NoStop}%
\bibitem [{\citenamefont {Kraus}\ and\ \citenamefont
  {Larsen}(2007)}]{Kraus:2006nb}%
  \BibitemOpen
  \bibfield  {author} {\bibinfo {author} {\bibfnamefont {P.}~\bibnamefont
  {Kraus}}\ and\ \bibinfo {author} {\bibfnamefont {F.}~\bibnamefont {Larsen}},\
  }\href {\doibase 10.1088/1126-6708/2007/01/002} {\bibfield  {journal}
  {\bibinfo  {journal} {JHEP}\ }\textbf {\bibinfo {volume} {01}},\ \bibinfo
  {pages} {002} (\bibinfo {year} {2007})},\ \Eprint
  {http://arxiv.org/abs/hep-th/0607138} {arXiv:hep-th/0607138 [hep-th]}
  \BibitemShut {NoStop}%
\bibitem [{\citenamefont {Mizoguchi}(1989)}]{Mizoguchi:1988vk}%
  \BibitemOpen
  \bibfield  {author} {\bibinfo {author} {\bibfnamefont {S.}~\bibnamefont
  {Mizoguchi}},\ }\href {\doibase 10.1016/0370-2693(89)91256-2} {\bibfield
  {journal} {\bibinfo  {journal} {Phys. Lett.}\ }\textbf {\bibinfo {volume}
  {B222}},\ \bibinfo {pages} {226} (\bibinfo {year} {1989})}\BibitemShut
  {NoStop}%
\bibitem [{\citenamefont {Kraus}\ and\ \citenamefont
  {Perlmutter}(2011)}]{Kraus:2011ds}%
  \BibitemOpen
  \bibfield  {author} {\bibinfo {author} {\bibfnamefont {P.}~\bibnamefont
  {Kraus}}\ and\ \bibinfo {author} {\bibfnamefont {E.}~\bibnamefont
  {Perlmutter}},\ }\href {\doibase 10.1007/JHEP11(2011)061} {\bibfield
  {journal} {\bibinfo  {journal} {JHEP}\ }\textbf {\bibinfo {volume} {11}},\
  \bibinfo {pages} {061} (\bibinfo {year} {2011})},\ \Eprint
  {http://arxiv.org/abs/1108.2567} {arXiv:1108.2567 [hep-th]} \BibitemShut
  {NoStop}%
\bibitem [{\citenamefont {Gaberdiel}\ and\ \citenamefont
  {Gopakumar}(2012)}]{Gaberdiel:2012ku}%
  \BibitemOpen
  \bibfield  {author} {\bibinfo {author} {\bibfnamefont {M.~R.}\ \bibnamefont
  {Gaberdiel}}\ and\ \bibinfo {author} {\bibfnamefont {R.}~\bibnamefont
  {Gopakumar}},\ }\href {\doibase 10.1007/JHEP07(2012)127} {\bibfield
  {journal} {\bibinfo  {journal} {JHEP}\ }\textbf {\bibinfo {volume} {07}},\
  \bibinfo {pages} {127} (\bibinfo {year} {2012})},\ \Eprint
  {http://arxiv.org/abs/1205.2472} {arXiv:1205.2472 [hep-th]} \BibitemShut
  {NoStop}%
\end{thebibliography}%



\end{document}